\documentclass[12pt,preprint]{aastex}

\newcommand{\msun}    {M_{\odot}}  
\newcommand{\yr}    {\rm yr}

\slugcomment{To appear in The Astrophysical Journal}

\shorttitle{A Model for the G31.41+0.31 Hot Core}
\shortauthors{Osorio et al.}

\begin{document}

\title{Collapsing Hot Molecular Cores: A Model for the Dust Spectrum
and Ammonia Line Emission of the G31.41+0.31 Hot Core}

\author{Mayra Osorio, Guillem Anglada}
\affil{Instituto de Astrof\'\i sica de Andaluc\'\i a, CSIC,
Camino Bajo de Hu\'etor 50, E-18008 Granada, Spain; osorio@iaa.es, 
guillem@iaa.es}

\author{Susana Lizano, Paola D'Alessio}
\affil{Centro de Radioastronom\'{\i}a y Astrof\'\i sica, UNAM, 
Ap. Postal 72-3, 58089 Morelia, Michoac\'an,  Mexico;
s.lizano@astrosmo.unam.mx, p.dalessio@astrosmo.unam.mx}

\begin{abstract}

We present a model aimed to reproduce the observed spectral energy 
distribution (SED) as well as the ammonia line emission of the G31.41+0.31 
hot core. The hot core is modeled as an infalling envelope onto a massive 
star that is undergoing an intense accretion phase. We assume an envelope 
with a density and velocity structure resulting from the 
dynamical collapse of a singular logatropic sphere. The stellar and 
envelope physical properties are determined by fitting the observed SED.
 From these physical conditions, the emerging ammonia line emission is
calculated and compared with subarcsecond resolution VLA data of the (4,4)  
transition taken from the literature. The only free parameter in this line
fitting is the ammonia abundance. The observed intensities of the main and
satellite ammonia (4,4) lines and their spatial distribution can be well
reproduced provided it is taken into account the steep increase of the
gas-phase ammonia abundance in the hotter ($> 100$ K), inner regions of
the core produced by the sublimation of icy mantles where ammonia
molecules are trapped. The model predictions for the (2,2), (4,4), and
(5,5) transitions, obtained with the same set of parameters, are also in
reasonably agreement, given the observational uncertainties, with the
single-dish spectra of the region available in the literature. The best
fit is obtained for a model with a central star of $\sim25~M_{\odot}$, a
mass accretion rate of $\sim3 \times 10^{-3}~M_{\odot}$~yr$^{-1}$, and a
total luminosity of $\sim 2 \times 10^{5}~L_{\odot}$. The outer radius of
the envelope is 30,000 AU, where kinetic temperatures as high as $\sim 40$
K are reached. The gas-phase ammonia abundance ranges from $\sim 2 \times
10^{-8}$ in the outer region to $\sim 3 \times 10^{-6}$ in the inner
region. To our knowledge, this is the first time that the dust and
molecular line data of a hot molecular core, including subarcsecond
resolution data that spatially resolve the structure of the core, have
been simultaneously explained by a detailed, physically self-consistent
model. This modeling shows that hot, massive protostars are able to excite
high excitation ammonia transitions up to the outer edge ($\sim$ 30,000
AU) of the large scale infalling envelopes.

\end{abstract}


\section{Introduction}
\label{introdu}

Hot molecular cores (hereafter HMCs) are small, dense, hot, and dark
molecular clumps, usually found in the proximity of ultracompact HII
regions (e.g., Kurtz et al. 2000). Unlike HII regions, these objects
present weak or undetectable free-free emission. The lack of free-free
emission has been interpreted as due to an intense mass accretion phase
that quenches the development of an HII region (e.g., Walmsley 1995).  
Therefore, these HMCs may be the precursors of HII regions, tracing the
earliest observable stage in the life of massive stars. A detailed
modeling of the emission of the HMCs is a necessary requirement to assess
their physical nature and their relationship with the birth of massive
protostars. Osorio, Lizano, \& D'Alessio (1999) showed that the spectral
energy distribution (SED) of the dust continuum emission of several HMCs
without detectable free-free emission is consistent with that of a massive
envelope collapsing onto a B type star, with an age less than $10^5$ yr,
and accreting mass at a rate of $10^{-4}$--$10^{-3}~\msun~\yr^{-1}$.  
These intense mass accretion rates are high enough to prevent the
development of an ionized region around the central star.  The physical
properties of the HMCs were inferred using the collapse of the Singular
Logatropic Sphere (SLS; McLaughlin \& Pudritz 1997) as a dynamical model
for the envelope. The SLS collapse models were chosen because these
envelopes are massive enough to reproduce the strong millimeter emission
observed, without requiring excessively high mass accretion rates, thus
preventing excessive emission in the near and mid infrared (see more 
details in Osorio et al. 1999).

One can further test Osorio's models, including the velocity field of the 
accreting envelopes, by comparing their expected molecular emission with 
observations of appropriate molecular tracers.  High-excitation ammonia 
transitions are well suited for this kind of comparison, since they trace 
the hot and warm dense gas in the HMCs, and can be observed with 
subarcsecond angular resolution using the Very Large Array (VLA).  Such 
observations are among the highest angular resolution molecular line data 
of HMCs that can be obtained at present, allowing to determine the 
variations of the line intensity across the sources.  Furthermore, the 
ammonia lines have hyperfine structure that is sensitive to the optical 
depth and permit an accurate measurement of the column density.

In order to do this test, we selected the HMC located $\sim 5''$ to the
southwest of the peak of the G31.41+0.31 ultracompact HII region.  
Hereafter, we will call this source G31 HMC.  This HMC is located at a
distance of 7.9 kpc (Cesaroni et al. 1998).  It is an interesting HMC
because it exhibits strong millimeter continuum emission, as well as
molecular emission of high excitation lines; it is associated with a group
of water masers, and with a bipolar outflow (Cesaroni et al. 1994, 1998;
Gibb, Wyrowski \& Mundy 2004). The presence of high excitation molecular
emission in this luminous source indicates that G31 HMC (together with a
few other objects;  Beuther \& Walsh 2008, Longmore et al. 2007) is one of
the hottest molecular cores discovered so far. The lack of strong
free-free centimeter emission (Araya et al. 2003) indicates that G31 HMC
could be in a phase of intense mass accretion, being unable to develop a
detectable HII region. Therefore, we consider G31 HMC as a good candidate
to test the basic hypothesis of Osorio's model, namely, that young massive
stars are formed inside HMCs with very high accretion rates. In this model
the onset of a detectable ultracompact HII region is quenched by the very
high mass accretion rate.

G31 HMC has a database of continuum flux density measurements covering a
wide range of wavelengths although, unfortunately, most of the data points
were obtained with relatively poor angular resolution and the SED is
poorly constrained for $\lambda \le 850~\mu$m. G31 HMC is associated with
line transitions from a variety of molecular species, such as HCO$^{+}$,
SiO (Maxia et al. 2001), $^{13}$CO (Olmi et al. 1996), CS (Anglada et al.
1996), H$_2$S, C$^{18}$O (Gibb, Mundy \& Wyrowski 2004), and CH$_3$CN
(Beltr\'an et al. 2005). Particularly interesting is the association of
G31 HMC with strong ammonia emission (Churchwell et al. 1990, Cesaroni et
al. 1992) whose (4,4) inversion transition has been observed with
subarcsecond angular resolution using the VLA (Cesaroni et al. 1998; see
their Figs. 2c, 4c, and 9c). This makes G31 HMC one of the few sources
where a high signal-to-noise ratio analysis of the spatial variation of
the ammonia emission along the core can be carried out. The high intensity
of the observed ammonia (4,4)  emission, the ratio of the main to
satellite lines close to unity, as well as the unusually large line widths
indicate extreme physical conditions that suggest that this core is
harboring an O type star. Because of this, we chose G31 HMC as a candidate
to test the molecular emission predictions of the Osorio's model for the
early stages of the formation of massive stars.

We note that VLA observations at 7 mm (Araya et al. 2003) reveal the 
presence of a binary system at the center of G31 HMC.  However, the model 
of Osorio et al. (1999) is spherically symmetric, with a single central 
source of luminosity. This assumption is valid as long as the binary 
separation ($\sim0\rlap.''2$) is much smaller than the observed size of 
the HMC ($\sim8''$). Likewise, deviations from spherical symmetry either 
by rotation (disks) or by an intrinsic elongation of the cloud are not 
considered by Osorio's model.  To test non-spherical models, one must 
include high angular resolution mid-IR data that would allow to properly 
constrain parameters such as the degree of flattening of the envelope, the 
centrifugal radius, or the effect of a disk and/or cavities caused by 
outflows. This kind of data has been obtained for other HMCs and is highly 
sensitive to the geometry of the cloud (e.g., De Buizer, Osorio \& Calvet 
2005). Unfortunately, such observations are not yet available for G31 HMC.

Observations of several molecular tracers reveal, however, the existence
of small velocity gradients in G31 HMC that have been interpreted as
suggestive of either outflow (Gibb, Wyrowski, \& Mundy 2004, Araya et al.  
2008) or rotation motions (Beltr\'an et al. 2005). Sensitive high angular
resolution observations are needed to clarify this issue. Despite the
possible presence of small velocity gradients, we expect the kinematics of
the core to be dominated by the overall infall motions of the envelope. In
this work we assume that the spherically symmetric models of Osorio et al.  
(1999) are adequate to study the dust continuum and molecular line
emission of G31 HMC.

The paper is organized as follows. In $\S 2$, we describe the modeling 
procedure, in $\S 3$ we present the observational data available from the 
literature, in $\S 4$ we model the dust and ammonia data and we compare 
our results with the observations, as well as with other modeling attempts 
of this source reported in the literature. Finally, in $\S 5$ we summarize 
our conclusions.

\section{Modeling \label{modelo}}

\subsection{General Procedure}

Following Osorio et al. (1999), we model a HMC as a spherically symmetric 
envelope of dust and gas that is collapsing onto a recently formed central 
massive star.  An accretion shock is formed at the stellar surface. Thus, 
the stellar radiation and the accretion luminosity provide the source of 
heating of the envelope.  The temperature of the envelope is a function of 
the distance to the center. The density, infall velocity, and turbulent 
velocity dispersion for each point of the envelope are obtained from the 
solution of the dynamical collapse of the SLS (McLaughlin \& Pudritz 
1997). The parameters of the SLS model are chosen by calculating the 
emerging dust emission, assuming a constant dust to gas mass ratio along 
the envelope, and fitting the observed SED.

Once obtained the physical conditions in the envelope from the fit to the 
observed SED, the excitation of the ammonia molecule is calculated and the 
radiative transfer is performed in order to obtain the emerging ammonia 
spectra for different lines of sight towards the envelope. In these 
calculations, the gas-phase ammonia abundance, that is a function of 
radius, is the only free parameter.  We consider two simple cases: uniform 
gas-phase abundance and uniform total ammonia abundance. In the latter 
case, the ratio of gas to solid phase ammonia as a function of radius is 
determined by calculating the balance between condensation and sublimation 
of molecules on dust grains.  The gas-phase ammonia abundance is 
constrained by comparing the ammonia model spectra with the observed 
spectra towards different positions.

\subsection {Physical Structure of the Envelope}
\label{structure}

For the collapsing envelope, we adopt the density, $\rho_{\rm SLS}(r)$,
infall velocity, $v_{\rm SLS}(r)$, and turbulent velocity dispersion,
$\sigma_{\rm SLS}(r)$, distributions resulting from the solution of the
self-similar collapse of the SLS (Adams, Lizano \& Shu 1987, unpublished
notes).  The SLS has a logatropic pressure, $P = P_0 \ln (\rho_{\rm
SLS}/\rho_0)$, where $P_0$ is a constant that sets the pressure scale, and
$\rho_0$ is an arbitrary reference density, introduced by Lizano \& Shu
(1989) to empirically take into account the observed turbulent motions in
molecular clouds. The collapse of the SLS has been studied by McLaughlin
\& Pudritz (1996, 1997; hereafter MP97), and it has been generalized by
Reid, Pudritz \& Wadsley (2002), and Sigalotti, de Felice \& Sira (2002)
to include three-dimensional hydrodynamic calculations of the collapse of
both singular and nonsingular logatropic spheres.

In the SLS collapse solution an expansion wave moves outward into a static
cloud and sets the gas into motion towards the central star. Both the
speed of the expansion wave and the mass accretion rate increase with
time. The radius of the expansion wave is given by
 \begin{equation}
 r_{\rm ew} = {1\over4}(2 \pi G P_0)^{1/2}\,t^2,
 \label{eq:rew}
 \end{equation}
 where $t$ is the time elapsed since the onset of collapse. 

Outside the radius of the expansion wave ($r>r_{\rm ew}$), the SLS 
envelope is static,
 \begin{equation}
v_{\rm SLS}=0,
 \label{eq:statvel}
 \end{equation}
 and the density is given by
 \begin{equation}
\rho_{\rm SLS}(r)=(P_0/2 \pi G)^{1/2}\,r^{-1}.
 \label{eq:statden}
 \end{equation}

Inside the expansion wave ($r<r_{\rm ew}$), 
 the infall velocity, whose direction is radial, is given by
\begin{equation}
 v_{\rm SLS}(r)={{1}\over{2}}(2 \pi G P_0)^{1/2} t\,u(x),
 \label{eq:logavel}
 \end{equation}
having a free-fall behavior for $r \ll r_{\rm ew}$. The density is
 \begin{equation}
 \rho_{\rm SLS}(r) =  {{2}\over{\pi^{1/2} G t^2}} \alpha(x),
 \label{eq:logaden}
 \end{equation}
 where $x = 4 (2 \pi G P_0)^{-1/2}\,r\,t^{-2}$ is the similarity variable,
and $u(x)$ and $\alpha(x)$ are non-dimensional functions. With this
normalization the expansion wave is located at $x=1$. The self-similar 
variable, and the density and velocity functions are related to those 
tabulated by McLaughlin \& Pudritz (1997) through $x=2^{5/2}
x_{\rm MP}$, $\alpha(x)= 2^{-3} \alpha_{\rm MP}(x_{\rm MP})$, and $u(x)=
2^{3/2} u_{\rm MP}(x_{\rm MP})$, where the subindex ``MP'' labels the 
McLaughlin \& Pudritz (1997) solution.

As noted by Osorio et al. (1999) the SLS collapse tends to produce massive
envelopes, since inside the radius of the expansion wave only 3\% of the
mass is in the central star, while 97\% is in the collapsing envelope.

The velocity dispersion inside the envelope due to turbulent 
motions is obtained assuming that Alfv\'en waves in the cloud induce fluid 
motions with velocity amplitudes, $\delta v_{\rm tur}$, of the order of 
the wave speed, $\delta v_{\rm tur}\simeq v_A= (dP/d\rho_{\rm 
SLS})^{1/2}=(P_0/\rho_{\rm SLS})^{1/2}$, where the magnitude of the 
magnetic field is $B=(4\pi P_0)^{1/2}$. For random polarizations and 
random orientations of the magnetic field in the cloud, the 
one-dimensional velocity dispersion in the line-of-sight direction is 
given by
 \begin{equation}
\sigma_{\rm SLS}(r) = \left(\frac{1}{3}\frac{P_0}{\rho_{\rm SLS}}\right)^{1/2},
\label{eq:dispersion}
\end{equation}
where the factor 1/3 comes from averaging over the solid angle the square
of the magnetic field projections along the line of sight. 

The physical structure of the SLS collapse is characterized by the 
constant $P_0$ and the time elapsed since the onset of collapse, $t$.
In terms of familiar variables, this elapsed time is given by
 \begin{equation}
t=\frac{4M_*}{\dot M},
 \end{equation}
where $\dot M$ is the mass accretion rate, and $M_*$ is the mass of the 
central star.
$P_0$ can be written as
 \begin{equation}
P_0=\frac {\dot M^{8/3}} {(2^{11} \pi^{3} G
m^2_0 M^{6}_*)^{1/3}},
\label{eq:p0}
 \end{equation}
 where $m_0=0.0302$ is the reduced (dimensionless) mass.
 From now on, we will use $M_*$ and $\dot M$ to characterize the
dynamical model. 

The total source of heating of the envelope, $L_{\rm tot}$, is assumed to 
be the sum of the stellar luminosity, $L_*$, and the accretion luminosity, 
$L_{\rm acc}=GM_*{\dot M}/R_*$, where $R_*$ is the stellar 
radius. A value of $R_*=1\times10^{12}$ cm has been adopted (see Osorio et 
al. 1999, Hosokawa \& Omukai 2008). The stellar luminosity is related to 
the stellar mass using the Schaller et al. (1992) evolutionary tracks. The 
temperature of the dust grains inside the envelope, $T(r)$, is 
self-consistently calculated from the total luminosity using the condition 
of radiative equilibrium for outer optically thin regions of the envelope, 
whereas for the inner optically thick regions the temperature is 
calculated from the standard diffusion approximation (see details in 
Osorio et al.  1999).

For temperatures $\ga60$ K and densities $\ga10^5$ cm$^{-3}$, the gas and 
dust are well coupled and are described by the same temperature (e.g., 
Sweitzer 1978; Doty et al. 2002). Our models fulfill this condition and, 
thus, the kinetic temperature of the gas, $T_k$, is taken to coincide with 
the temperature obtained from the dust calculations, $T_k(r)=T(r)$.

Assuming that the dust grains shield the molecules from the intense
central radiation field, the dust destruction front, at the dust
sublimation temperature ($\sim 1200$ K), determines the inner radius,
$R_{\rm dust}$, of the dust and molecular envelope. The external radius,
$R_{\rm ext}$, of the envelope is inferred from the observations.

\subsection{Dust Continuum Emission}
\label{modeldust}

\begin{figure}
\vspace{9.0cm}
\includegraphics{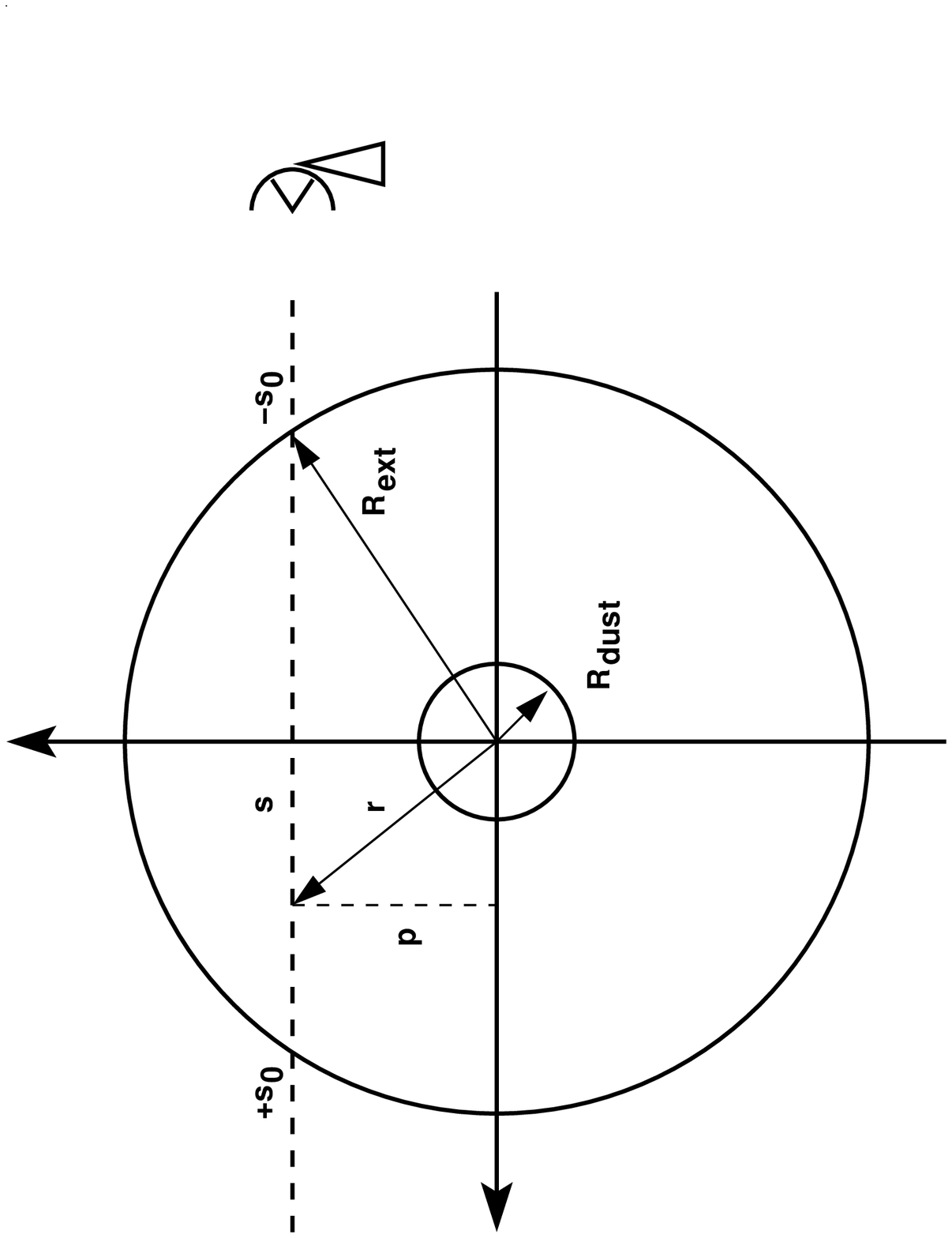}
\caption{
 Geometry of the HMC. The plane containing the center of the core and the 
line of sight is shown. $R_{\rm dust}$ and $R_{\rm ext}$ are the inner and
outer radii of the envelope, respectively, $p$ is the impact parameter of
the line of sight, $r$ is the radius of a given point, $s$ its coordinate 
along the line of sight, and $-s_0$, $+s_0$ are the coordinates of the 
envelope edges.}
 \label{fig:cartoon}
 \end{figure}

To solve the radiative transfer equation for the dust emission we have the 
geometry depicted in Figure~\ref{fig:cartoon}. Given the spherical 
symmetry of the envelope, the emergent specific intensity, $I_\nu$, for a 
given line of sight is only a function of the impact parameter, $p$, the 
distance between the line of sight and the center of the envelope. 
Assuming that the source function of the thermal dust emission is the 
Planck function evaluated at the dust temperature, $B_\nu(T)$, we have the 
following set of equations that allow us to obtain the emerging intensity 
as a function of frequency and impact parameter,
 \begin{equation}
 I_\nu(\nu,p)=I_{\rm bg}(\nu)~{e}^{-\tau} + \int^{\tau}_0
B_{\nu}(T)~{e}^{-\tau'}~d\tau',
 \label{eq:tralin1p}
 \end{equation} 
 $$\tau'(\nu,p,s')=\int^{s'}_{-s_0}\kappa(\nu)\,\rho_{\rm SLS}(s,p)~ds,$$ 
 and
 $$\tau(\nu,p)=\int^{+s_0}_{-s_0} \kappa(\nu)\,\rho_{\rm SLS}(s,p)~ds,$$ 
 where $I_{\rm bg}(\nu)$ is the background intensity, $\kappa(\nu)$ is the
monochromatic dust absorption coefficient per unit mass, $s$ is the
coordinate along the line of sight ($r^2=p^2+s^2$), positive in the
direction away from the observer, and whose origin is in the plane
perpendicular to the line of sight that contains the center of the
envelope (see Fig.~\ref{fig:cartoon}).  The integration limits $-s_0$ and
$+s_0$ correspond to the edges of the envelope nearest and farthest to the
observer, respectively, being $s_0=\sqrt{R^2_{\rm ext}-p^2}$.  In the
absence of a strong background source, $I_{\rm bg}$ will be taken as the
cosmological background, $I_{\rm bg}=B_{\nu}(T_{\rm bg})$, where $T_{\rm
bg}= 2.7$ K.

Following Osorio et al. (1999), the dust absorption coefficient,
$\kappa(\nu)$, at high frequencies ($\nu>1500$ GHz) is obtained from
D'Alessio (1996), who considered spherical (Mie) particles of graphites,
silicates, iron, and water ice compounds, with the standard grain-size
distribution of the interstellar medium, $N(a) \propto a^{-3.5}$ (Mathis,
Rumpl, \& Nordsieck 1977), with a minimum radius of 0.005 $\mu$m and a
maximum radius of 0.3 $\mu$m. The abundance and optical constants of the
compounds were taken from Draine \& Lee (1984), Draine (1987), and Warren
(1984). The value of the absorption coefficient at 1500 GHz obtained for
this dust composition and size distribution can be extrapolated to lower
frequencies in the standard power-law form, resulting $\kappa(\nu) = 0.06
(\nu / {\rm 1500~GHz})^\beta$ cm$^{2}$ g$^{-1}$ for $\nu \le 1500$ GHz.
The value of the index $\beta$ is adopted so that it is consistent with
the slope of the optically thin millimeter emission of the observed SED,
where $S_\nu \propto \nu^{(2+\beta)}$.  We assumed a constant dust-to-gas
mass ratio of 0.01. Depending of the wavelength, the D'Alessio (1996)  
opacities can be a factor of up to 2-3 times lower than those of Ossenkopf
\& Henning (1994). Both opacity models have been successfully used to
reproduce the SED of other high mass star forming cores (Osorio et al
1999, van der Tak et al 2000).

Once the emergent intensity as a function of the impact parameter is
obtained, the total flux density of the source at a given frequency is
calculated as
 \begin{equation} F_\nu(\nu)={{2\pi}\over{D^2}}\int^{R_{\rm ext}}_0
I_\nu(\nu,p)\,p\,dp,
 \end{equation}
 where $D$ is the distance of the observer to the source.  Finally, the
resulting model SED is calculated and compared with the observed SED.  A
grid of models is run with different values of the free parameters $\dot
M$ and $M_*$ (or, equivalently, $L_*$). In this way, the model (or models)
that are consistent with the observed SED are determined. Note that in
case the SED is not well constrained, more than one model can be
consistent with the data. Further details on the method are described by
Osorio et al. (1999).

\subsection {Ammonia Line Emission}
\label{am1}

In this section, we describe the calculation of the emerging intensity of
the ammonia inversion transitions arising from a spherically symmetric
envelope with the physical properties obtained from the dust modeling
using the SLS solution.

The physics of the ammonia molecule is studied in detail in Townes \&
Slawlow (1975), Ho (1975), and Ho \& Townes (1983). Here we only summarize
some basic information that may be relevant to our discussion. Because
ammonia is a symmetric top molecule, its rotational energy levels are
described by two quantum numbers, namely, the total angular momentum, $J$,
and its projection on the symmetry axis, $K$ ($K\le J$). Rotational levels
with $J=K$ are more populated and are called metastable, while rotational
levels with $J\ne K$ are short-lived and are called non-metastable. All
rotational levels (except those with $K=0$) are split into inversion
doublets. The transitions between these inversion doublets are called
inversion transitions and occur at wavelengths of $\sim$1.3 cm. 

As shown in Appendix~\ref{apenTex}, the excitation temperature of the
ammonia inversion transitions can be obtained from a two-level model. For
a given point of coordinate $s$ along a line of sight of impact parameter
$p$, we can estimate the excitation temperature, $T_{\rm ex}(s,p)$, from
equation~(\ref{eq:dosni2}), taking the total number density of gas
molecules as $n(r)=\rho_{\rm SLS}(r)/(\mu_m\,m_{\rm H}$), where
$\mu_m=2.3$ is the mean molecular weight in the envelope and $m_{\rm H}$
is the hydrogen mass. In the calculation of $T_{\rm ex}$ we will
approximate the intensity of the local radiation field by the background
intensity ($I_{r}(s,p)\simeq I_{\rm bg}$). With this approximation,
equation~(\ref{eq:dosni2}) gives a lower limit for the excitation
temperature. However, as shown in Appendix~\ref{apenTex}, for the physical
conditions of HMCs the values obtained in this way are very close to the
value of the kinetic temperature, which, for thermal lines, is an upper
limit for the excitation temperature. Therefore, this indicates that
taking $I_{r}=I_{\rm bg}$ is a very good approximation for the calculation
of the excitation temperature in HMCs.  Note that, with this
approximation, the resulting excitation temperature will become a
spherically symmetric function, $T_{\rm ex}(r)$.

The spectra of the inversion transitions present a hyperfine structure due
to the interaction of the electric quadrupole momentum of the nitrogen
nucleus with the electric field of the electrons. This effect splits the
line into five components, the central main line, and two pairs of
satellite lines, separated $\sim$1-2 MHz, symmetrically placed about the
main line. These hyperfine components are further split by weaker magnetic
spin interactions. Since the separation between the magnetic hyperfine
components is only $\sim$10-40 kHz (0.1-0.5 km s$^{-1}$), they are
distinguishable only in very high spectral resolution observations of
regions with very low velocity dispersion. For this reason, we neglect the
magnetic hyperfine structure, and we restrict our analysis to only the
five electric quadrupole hyperfine components.

As shown in Appendix~\ref{apenkappa} (eq.~[\ref{eq:opafinal}]), the
absorption coefficient of the ammonia $(J,K)$ inversion transition, at a
given point of coordinate $s$ along a line of sight of impact parameter
$p$ (see Fig.~\ref{fig:cartoon}), as a function of the observed LSR
velocity, $V$, is given by
 \begin{eqnarray}
 \kappa(V,s,p) & = & {
 \left(1 \over 128\pi^3\right)^{1/2} {c^3 A_{ul} \over \nu_0^3}
 {{\rho_{\rm SLS}(r) X_{\rm NH_3}(r)} \over {1.2 \mu_m m_{\rm H} 
\sigma_V(r)}}
 {e^{h \nu_0/k T_{\rm ex}(r)}-1 \over e^{h \nu_0/k T_{\rm ex}(r)}+1}
 } \nonumber \\
 & & \times {g_{JK} e^{-E_{JK}/k T_{\rm rot}(r)} \over Q(T_{\rm rot}(r))} 
 \sum_{i=1}^5 x_i e^{-\left(V - V_i - V_{c} -
 v_{\rm SLS}(r)\,s/r\right)^2/2\sigma_V^2(r)},
 \label{eq:opafinal2}
 \end{eqnarray}
 where the frequency $\nu_0$ corresponds to that of the main line of the
inversion transition (given in Table~\ref{tbl:inver}), $X_{\rm NH_3}(r)$ 
is the local gas-phase ammonia abundance relative to H$_2$, $\sigma_V(r)$ 
is the local dispersion of the distribution of line-of-sight velocities 
due to turbulent and thermal motions, $T_{\rm rot}(r)$ is the local 
rotational temperature, $g_{JK}$ is the statistical weight of the $(J,K)$ 
rotational level (given by equation~[\ref{eq:pesos}]), $E_{JK}$ its 
rotational energy (given by equation~[\ref{eq:energy}]), $Q(T_{\rm rot})$ 
is the partition function (given by equation~[\ref{eq:particion}]), $x_i$ 
is the relative strength of the $i$th hyperfine component ($\sum_{i=1}^5 
x_i =1$) and $V_i$ its Doppler velocity shift with respect to the main 
line (both given in Table~\ref{tab:propnh3}), $V_{c}$ is the LSR velocity 
of the ambient cloud, and the term $v_{\rm SLS}(r)\,s/r = V_s$ is the 
line-of-sight component of the infall velocity of the envelope. In writing 
equation~(\ref{eq:opafinal2}) it has been assumed that $n=1.2 n_{\rm 
H_2}$, corresponding to assume a 10\% He abundance. Note that the 
excitation temperature, $T_{\rm ex}$, and the Einstein spontaneous 
emission coefficient, $A_{ul}$, are those of the entire inversion 
transition, unsplit by hyperfine interactions (eq.~[\ref{eq:dosni2}] and 
Table~\ref{tbl:inver}, respectively). Due to the dipolar selection rules 
on the rotational transitions, the populations of the metastable levels 
are mainly determined by collisions; therefore, $T_{\rm rot}$, that is 
defined by the relative populations of the rotational levels is expected 
to be similar to the kinetic temperature. In fact, $T_{\rm rot}$ is a 
lower limit of $T_k$, and its value is different for each pair of 
rotational levels considered (see Appendix~\ref{apenkappa}). However, the 
agreement between $T_{\rm rot}$ and $T_k$ is expected to improve for high 
metastable levels, due to the increased excitation involved, and for high 
densities, where all temperatures tend to thermalize to $T_k$ (e.g., 
Walmsley \& Ungerechts 1983; Danby et al. 1988; Wilson et al. 2006). This 
is the case for G31 HMC (Osorio 2000), and hereafter we will take $T_{\rm 
rot}$=$T_k$.

In the above equation, the dispersion of the line-of-sight velocity
distribution results from the contribution of the macroscopic (turbulent)
and microscopic (thermal) components,
 \begin{equation}
 \sigma_V(r)=\sqrt{\sigma^2_{\rm SLS}(r)+ \sigma^2_{\rm th}(r)},
 \label{eq:DeltaV}
 \end{equation}
 where $\sigma_{\rm SLS}(r)$ is given by equation~(\ref{eq:dispersion}) 
and $\sigma_{\rm th}(r)$ is given by
 \begin{equation}
 \sigma_{\rm th}(r)={{k T_k(r)} \overwithdelims () {m_{\rm NH_3}}}^{1/2},
 \label{eq:Deltath}
 \end{equation}
where $m_{\rm NH_3}$ is the mass of the ammonia molecule.

The only free parameter in equation~(\ref{eq:opafinal2}) is the gas-phase
ammonia abundance, $X_{\rm NH_3}(r)$.  The smallest values of the ammonia
abundance reported in the literature are found in the coolest cores of
low-mass star forming regions, with typical abundances of $X_{\rm
NH_3}=10^{-8}$-$10^{-7}$ (e.g., Herbst \& Klemperer 1973, Ungerechts et
al. 1980, Estalella et al. 1993), and some chemical models predict
abundances as low as $X_{\rm NH_3}=10^{-10}$-$10^{-9}$ (Le Bourlot et al.
1993). The highest values, obtained in very hot regions of massive star
formation (e.g., Millar 1997, Walmsley 1997, Ohishi 1997), are $X_{\rm
NH_3}=10^{-6}$-$10^{-5}$.

Neglecting chemical effects, we consider the two simplest possibilities
for the function, $X_{\rm NH_3}(r)$, that describes the gas-phase ammonia
abundance inside the envelope. First, we consider the case in which
$X_{\rm NH_3}(r)$ has a constant value across the envelope. As a second
possibility, we consider the case in which the total (solid+gas) ammonia
abundance remains constant across the envelope but the gas-phase abundance
increases towards the center because of the temperature gradient.
 Variations in the gas-phase molecular abundances have already been
explored by, e.g., van der Tak et al. (2000), and Boonman et al. (2003) to
explain the CH$_3$OH, HCN, and H$_2$O emission in some regions of massive
star formation. We assume a minimum gas-phase ammonia abundance, $X_{\rm
min}$, in the outer, cooler parts of the core, where the bulk of the
ammonia molecules are frozen in grain mantles. The ammonia molecules are
released from the grain mantles to the gas phase, reaching a maximum value
of the gas-phase ammonia abundance, $X_{\rm max}$, in the inner, hotter
regions of the core, where it is assumed that all the ammonia molecules
are in the gas phase. Under this simple approximation, the gas-phase
ammonia abundance as a function of the distance to the center of the
core will be given by
 \begin{equation}
 X_{\rm NH_3}(r)= {{X_{\rm max}-X_{\rm min}} \over {1 + \eta(n(r),T(r))}}
 + X_{\rm min},
 \label{eq:escalon} 
 \end{equation}
 where $\eta(n,T)$ is the ratio between the solid and gas phases, obtained
from the balance between the condensation and sublimation of molecules on 
dust grains following the thermal equilibrium equation of Sandford \& 
Allamandola (1993) (see Appendix~\ref{apeneta}). Two possible cases are 
considered. First, it is assumed that the ammonia molecules are directly 
frozen in grain mantles, with $\eta(n,T)$ described by 
equation~(\ref{eq:etanh3}); in this case, the ammonia molecules are 
released to the gas phase at the ammonia sublimation temperature, which is 
of $\sim$60 K for the range of densities typical in HMCs (see 
Fig.~\ref{fig:sketch}). As a second possibility, it is assumed that the 
ammonia molecules are trapped in water ice, being released to the gas 
phase only after sublimation of water molecules, that occurs at a 
temperature of $\sim$100 K as described by equation~(\ref{eq:etah2o}) (see 
Fig.~\ref{fig:sketch}).

As in the case of the continuum emission ($\S$~\ref{modeldust}), we adopt 
the geometry of Figure~\ref{fig:cartoon}, and the emergent specific intensity, 
$I_{\nu}$, will be given by
  \begin{eqnarray}
 I_{\nu}(V,p) & = & I_{\rm bg}~{e}^{-\tau} + \int^{\tau}_0 
B_{\nu}(T_{\rm
ex}(s,p))~{e}^{-\tau'}~d\tau', \nonumber \\
 & & \tau'(V,p,s')=\int^{s'}_{-s_0} \kappa(V,s,p)~ds, \nonumber \\
 & & \tau(V,p)=\int^{+s_0}_{-s_0} \kappa(V,s,p)~ds,
 \label{eq:tralin1}
 \end{eqnarray}
 where $T_{\rm ex}$ is given by equation~(\ref{eq:dosni2}), and
$\kappa(V,s)$ is given by equation~(\ref{eq:opafinal2}). As for the dust
calculations, in the absence of a strong background source, the background
intensity will be taken as the cosmological background, $I_{\rm
bg}=B_{\nu}(T_{\rm bg})$, where $T_{\rm bg}= 2.7$ K. The geometrical
parameters are the same as in equation~(\ref{eq:tralin1p}).

Finally, in order to make an accurate comparison with the observations,
the emergent intensity is convolved (in 2-D) with a Gaussian with a FWHM
equal to that of the observing beam, and a velocity smoothing is applied
to reproduce the spectral resolution of the spectrometer. Note that
$X_{\rm min}$ and $X_{\rm max}$ are the only parameters to be fitted by
the molecular line modeling.

\section{The Data \label{data}}

Numerous observations of the G31.41+0.31 region are reported in the
literature. The main results of the continuum flux density measurements in
the wavelength range $\sim 10~\mu$m to 1.3 cm are summarized in
Table~\ref{datos}. Most of these observations did not have the angular
resolution required to separate the emission of the HMC from that of the
G31.41+0.31 UCHII region, whose emission peak is only $\sim 5''$ apart.  
Thus, it is likely that in the data obtained with angular resolution
$>5''$ the dust continuum emission of the HMC is contaminated by emission
originated in the nearby UCHII. Therefore, we have considered these data
as upper limits. In addition, we note that the flux density at 1.3 cm is
likely dominated by free-free emission. Then, we expect the dust thermal
emission at this wavelength to be well below the upper limit set by the
observations, and therefore, in our modeling, we did not attempt to
reproduce the emission at 1.3 cm and longer wavelengths.  Finally, we
considered the 7 mm data point as a lower limit since it is likely that a
fraction of the flux density is missed in the very high angular resolution
interferometric VLA observations at this wavelength.

Observations of several line transitions from a variety of molecular
species have been carried out towards G31 HMC.  Among them, the VLA
B-array observations of the NH$_3$ (4,4) inversion transition (VLA project
code AH0483) reported by Cesaroni et al.  (1998) (combined with C-array
data from Cesaroni et al.  1994) constitute one of the best sets of high
angular resolution data towards G31 HMC and objects of the same kind
published so far. These combined data have an angular resolution of
$0\farcs63$ (corresponding to a spatial scale of 5000 AU at the distance
of 7.9 kpc of the source), high enough to spatially resolve the structure
of the core, revealing its extreme physical conditions and their variation
inside the core. Thus, we used this dataset to carry out the main part of
our spectral line data analysis.

Figure 9c of Cesaroni et al. (1998) shows the ammonia (4,4) spectra as a 
function of the impact parameter with respect to the center of G31 HMC. As 
can be seen in the figure, the brightness temperature is very high, 
reaching a value of approximately 100 K for the line-of-sight towards the 
center of the core, indicating high values of the kinetic temperature.  
The optical depth is also high, as indicated by the fact that the 
satellite lines, whose opacity is $\sim$1/60 that of the main line, reach 
an intensity similar to that of the main line towards the center of the 
core. Also, lines are unusually broad, resulting in the blending of 
contiguous satellite lines, indicative of large turbulent motions. This 
figure also indicates that the physical conditions in the core change as a 
function of the distance to the center, as shown by the variation of the 
line intensity and ratio between main line and satellites for different 
impact parameters.

In addition to the subarcsecond resolution ammonia data of Cesaroni et al.
(1998), Churchwell et al. (1990) and Cesaroni et al. (1992) report a
multitransitional ammonia study of G31 HMC carried out with the Effelsberg
100 m telescope. These observations have an angular resolution of $40''$.  
Although the study lacks the angular resolution to be sensitive to
variations of the line parameters across the core, providing only the
integrated ammonia emission of the overall core, these data are useful to
further test the model predictions derived from the analysis of the dust
and high angular resolution ammonia (4,4) line data.

In the next section we present a simultaneous analysis of the dust
continuum and ammonia line data that will provide a powerful diagnostic to
constrain the physical conditions inside G31 HMC.

\section{Comparison of the Model with the Observations}

\subsection{Dust Continuum Emission}
\label{compardust}

The observed SED of G31 HMC is characterized by strong millimeter emission
suggesting the presence of either a hot or a dense envelope (or a
combination of both). Therefore, ``a priori'' we expect the models that
are able to fit the observed SED to be characterized by a high mass
accretion rate. Bearing this in mind, we ran a grid of models covering a
relatively wide range of values of the free parameters $M_*$ and $\dot M$.
The mass and luminosity of the central star have been related using the
evolutionary tracks of Schaller et al. (1992) for a solar metallicity. We
tested the same values of the mass given in the tables of Schaller et al.
in the range $1~M_{\odot} \le M_* \le 40~M_{\odot}$ (corresponding to
$0.7~L_{\odot} \le L_* \le 2 \times10^5~L_{\odot}$), and for the mass
accretion rate we explored the range $10^{-5}~M_{\odot}~{\rm yr^{-1}} \le
{\dot M} \le 10^{-2}~M_{\odot}~{\rm yr^{-1}}$. We set the value of $\beta$
so that the slope of the model SED was consistent with that of the
observed SED in the millimeter wavelength range, resulting in values of
$\beta$ in the range 1-1.2.
  We set the external radius to $R_{\rm ext}$ = 30,000~AU, a value
inferred from the observed images of the source (e.g., Cesaroni et al.
1994a, Maxia et al.  2001) and required to reproduce the large flux
densities observed in the millimeter range.

Because of the incompleteness of the observed SED of G31 HMC, there is
more than one model that can fit the available observational data. In
Table~\ref{tbl5} we give the parameters of a set of models (corresponding
to values of the stellar mass listed in the tables of Schaller et al.  
1992) that are consistent with the observed SED. The higher luminosity
(and hotter envelope) model consistent with the observed SED has a central
star with a mass of $M_* = 25~M_\odot$ and an accretion rate of $\dot
M = 2.6 \times 10^{-3}~M_\odot~{\rm yr}^{-1}$.  Models with $M_* >
25~M_\odot$ produce far- and mid-IR flux densities that exceed the
observational constraints.  Lower luminosity models ($M_* < 25~M_\odot$)
that are consistent with the SED data have colder envelopes, but they have
a higher density (see the values of the temperature and density at
$r=1000$ AU listed in the table) in order to account for the strong
millimeter emission observed in G31 HMC, resulting in higher values for
the mass of the envelope. Note that a low value of $M_*$ does not imply
that the source is a low-mass protostar since in Table~\ref{tbl5} the
models with the smaller central mass are also the younger sources, meaning
simply that the time elapsed since the onset of collapse, $t$, is still
too short to accumulate a massive central object. In fact, as can be seen
in the table, these models are associated with the more massive envelopes
and would eventually form the higher mass stars.

The value of the inner radius of the envelope, $R_{\rm dust}$, is $\sim
156$ AU for model I ($M_* = 25~M_\odot$) and gets smaller as the mass
of the central star decreases.  We note that in all these models the
radius of the expansion wave, $r_{\rm ew}$, is smaller than the radius of
the core, $R_{\rm ext}$;  therefore, the outermost regions of these
envelopes are static. Also, the infall accretion rates in these models are
so high that the accretion luminosity is the most important source of
energy. For a given value of $M_*$, the observational uncertainties in the
SED data points result in a variation of $\sim 10\%$ in the possible
values of $\dot M$ that are consistent with the data. This results also in
a formal uncertainty of the order of 10\% in the derived parameters listed
in Table~\ref{tbl5} for a given value of $M_*$.

Figure~\ref{fields} shows the physical structure of the SLS envelopes
corresponding to the models listed in Table~\ref{tbl5}. As the figure
illustrates, model I has the higher values of the velocity and
temperature, and the lower values of the density and velocity dispersion.
Turbulent motions are the most important contributors to the local
velocity dispersion at large radii of the envelope, reaching values of
$\sigma_{\rm SLS} > 4$~km~s$^{-1}$ in the outermost regions.

\subsection{Ammonia (4,4) Line Emission}

Here we obtain the NH$_3$ emission for the models that are consistent with
the observed SED of G31 HMC to further test their physical properties by
comparing their emission with the VLA observations of the NH$_3$(4,4)
inversion transition reported by Cesaroni et al. (1998).  In order to
reproduce the setup of the Cesaroni et al.  observations, the emerging
ammonia intensity as a function of the LSR velocity, $V$, is calculated
for each line of sight through the core and convolved with a Gaussian beam
of HPBW = $0\farcs63$. The spectra have been converted into a brightness
temperature scale using the relation $T_{B}=2.1 (F_\nu / \rm
mJy~beam^{-1})(HPBW/ arcsec)^{-2}$. A velocity range of $\sim$ 150 km
s$^{-1}$ is covered, and Hanning smoothing is applied to obtain an
effective spectral resolution of 4.86 km s$^{-1}$.  A value of $V_c=97.4$
km s$^{-1}$ has been adopted for the LSR velocity of the ambient cloud.
Figure 9c of Cesaroni et al. (1998) shows the spectra obtained by
averaging the observed ammonia emission over circular annuli around the
position of the core center.  Therefore, in order to make more accurate
the comparison of the model results with the observations, the spectra
have been calculated for the impact parameters $p$ = 0, 5000, 10000, and
15000 AU, corresponding to the values sampled in this figure.

\subsubsection{Constant Gas-Phase Ammonia Abundance}

We first ran a grid of cases with different values of the gas-phase
ammonia abundance, assumed to be constant along the envelope. We tested a
wide range of values, from $X_{\rm NH_3}=10^{-9}$ to $X_{\rm
NH_3}=10^{-5}$, for all the models resulting from the fit to the observed
SED (discussed in $\S$~\ref{compardust}).  After running all these cases,
we conclude that none of these models is able to reproduce the observed
spectra.

Figure \ref{fig:grid1} shows the synthetic spectra obtained for model I at
different impact parameters and for different ammonia abundances. The
spectra observed by Cesaroni et al.  (1998) are also shown for comparison.
This figure shows that model I cannot reproduce the observed behavior of
the ammonia emission as a function of impact parameter. For a low ammonia
abundance (left column), the main line reaches the observed brightness
temperature of $\sim 100$ K in the $p=0$ panel;  however, the satellite
lines are too weak in all the panels. Increasing the ammonia abundance
(middle and right columns) produces an increase of the intensity of the
satellite lines, that reach a value close to the observed one ($\sim 100$
K) in the $p$ = 0 AU panel of the right column.  Nevertheless, in this
panel the main line reaches a brightness temperature of only $\sim 40$ K,
which is much smaller than the observed value. This occurs because the
main line is very optically thick (reaching an opacity of $\tau_{\rm
main}\simeq 350$) and, thus, the brightness temperature must be roughly
equal to the kinetic temperature of the outer parts of the envelope model,
which are cold. Note that the brightness temperature of the satellite
lines is higher since their opacity is 60 times lower, and they can reach
the inner parts of the envelope where the temperatures are higher.

Models II to XI are colder than model I (see Fig.~\ref{fields}) and 
therefore, they are worse in reproducing the observed behavior of the 
ammonia emission. As for model I, for a low abundance the satellite lines 
are too weak in all the panels. The intensity of the satellite lines 
increases with the ammonia abundance but, as in model I, for high ammonia 
abundances the main line becomes very optically thick in the $p$ = 0 AU 
panels ($\tau_{\rm main} > 380$) and only traces cold material from the 
outer envelope. Since models II to XI are colder than model I, the line 
intensities are lower and remain well below the observed values.

Clearly, there is no hope to reproduce the observations with constant
values of the ammonia abundance. A high optical depth is required to
reproduce the intense satellite lines, but this high opacity prevents that
the main line is formed in the inner, hottest regions resulting in a main
line brightness temperature too low. Nevertheless, the analysis of
constant abundance suggests the alternative of having a low abundance in
the outer envelope and a high abundance in the inner parts.  In this way,
the colder, outer parts of the envelope would have a smaller contribution
to the observed emission, that would instead trace inner, higher
temperatures. One expects that a compromise could be reached, in which the
ammonia abundance in the outer parts of the envelope is low enough for the
main line not to be saturated until reaching the inner, hotter regions,
where the abundance may be high enough for even the satellite lines to
become optically thick. In addition, having a low abundance in the outer
parts of the envelope would prevent the main line to become saturated in
the outer panels, helping to explain the observed decrease of the
intensity of the main line as the impact parameter increases.

\subsubsection{Variable Gas-Phase Ammonia Abundance}
\label{comparvar}

The rationale for a variable ammonia abundance is provided by the
calculation of the sublimation of molecules from dust grain mantles (see
Appendix~\ref{apeneta}).  Given the temperature gradient in the envelope,
the outer parts have temperatures below that of sublimation of grain
mantles and it is expected that most of the ammonia molecules are trapped
there. The ammonia gas-phase abundance should increase drastically in the
inner regions of the envelope, where the kinetic temperature reaches the
sublimation temperature of the mantles and the ammonia molecules are
released to the gas phase. Therefore, inclusion of the process of
sublimation of ammonia molecules from grain mantles should provide a more
realistic description of the abundance distribution required in the
calculation of the ammonia emission.

In order to try to fit the observed ammonia emission using a variable
gas-phase ammonia abundance, we first ran a grid of cases with the ammonia
abundance calculated using equation~(\ref{eq:escalon}), and with the
ratio, $\eta$, between the solid and gas phases given by
equation~(\ref{eq:etanh3}) of Appendix~\ref{apeneta} (that corresponds to
direct sublimation of NH$_3$ ices).  This procedure assumes that most of
the ammonia molecules are frozen into grain mantles in the outer parts of
the envelope, being released to the gas phase at the ammonia sublimation
temperature (about 60 K; see Appendix~\ref{apeneta} and
Fig.~\ref{fig:sketch}). In this way, the gas-phase ammonia abundance
behaves like a step function, with a minimum value, $X_{\rm min}$, in the
external parts of the envelope (where the temperature remains below the
ammonia sublimation temperature), and increasing steeply to its maximum
value, $X_{\rm max}$, for points inside the radius where the ammonia
sublimation temperature is reached.

 We have explored values for the outer gas-phase NH$_3$ abundance in the
range $10^{-10} \la X_{\rm min} \la 5\times 10^{-7}$, and in the range
$10^{-8} \la X_{\rm max} \la 10^{-4}$ for the inner gas-phase NH$_3$
abundance.  However, none of the models resulting from the fit to the
observed SED ($\S$~\ref{compardust}) is able to reproduce reasonably well
the NH$_3$(4,4) observations. The main problem is that the maximum
brightness temperatures of the main line obtained with this kind of
ammonia abundance distribution are too low ($\sim 60$ K), to account for
the peak brightness temperatures of $\sim 100$ K observed in G31 HMC.  
Apparently, this occurs because the main line becomes optically thick in
regions with temperatures of the order of 60 K, the temperature at which
the bulk of ammonia molecules that were frozen in grain mantles are
released to the gas phase.  Thus, the difficulties to fit simultaneously
the main and satellite lines are similar to those of the case of a
constant gas-phase ammonia abundance.

As already mentioned, an alternative scenario is that ammonia molecules
are mixed with water ice in grain mantles (as suggested, e.g., by A'Hearn
et al.  1987 for comets, and by Brown et al. 1988 and Osorio 2000 for
HMCs). In this case, ammonia molecules will be trapped in the grain
mantles until temperatures high enough for sublimation of water molecules
are reached. Therefore, the gas-phase ammonia abundance will be described
by equation~(\ref{eq:escalon}), where $\eta$ corresponds now to the water
sublimation and is given by equation~(\ref{eq:etah2o}) of
Appendix~\ref{apeneta}. As illustrated in Figure~\ref{fig:sketch}, the
transition from low to high gas-phase ammonia abundance occurs at a higher
temperature ($\sim 100$ K) than in the case of pure ammonia sublimation.  
As a result, optically thick lines will trace hotter regions of the
envelope, resulting in higher brightness temperatures.

As in the case of pure ammonia sublimation we ran a grid of cases with 
values for the outer gas-phase NH$_3$ abundance in the range $10^{-10} \la 
X_{\rm min} \la 5\times 10^{-7}$, and in the range $10^{-8} \la X_{\rm 
max} \la 10^{-4}$ for the inner gas-phase NH$_3$ abundance. We found that 
models II to XI cannot fit the NH$_3$ observations. The
envelopes of the lower luminosity models ($M_*\le 15~M_\odot$) are 
apparently too cold to fit the observed spectra. For these models the 
sublimation temperature of 100 K is reached at a radius $< 5000$ AU (see 
Fig.~\ref{fields}), and therefore the spectra calculated for the outer 
impact parameters ($p \ge 5000$ AU) behave very much like a case of 
constant gas-phase ammonia abundance equal to $X_{\rm min}$ (since the 
lines of sight for $p>5000$ AU only cross the outer regions of the 
envelope, whose temperatures remain below the sublimation temperature).  
Since models II to XI could not reproduce the observed spectra with a 
constant abundance, it is not expected that these models can fit the 
spectra in the outer panels, even with a variable gas-phase ammonia 
abundance. In summary, models with $M_*\le 15~M_\odot$ are too cold to 
reproduce the observed properties of the NH$_3$(4,4)  spectra, even with a 
variable gas-phase NH$_3$ abundance.

In contrast, we have been able to obtain a fairly good fit to the observed
spectra using model I ($M_*=25~M_\odot$).  In
Figure~\ref{fig:gridvariable} (central column) we show our best fit
spectra, obtained using this model with a value of $X_{\rm
min}=2\times10^{-8}$ and a value of $X_{\rm max}=3\times10^{-6}$. This set
of parameters reproduces quite well the observed spectra.  It reproduces
the observed intensity ($\sim 110$ K) of the main line towards the center
of the core ($p=0$), as well as the variation of its intensity as $p$
increases. It also reproduces the observed intensity of the satellite
lines in the $p=0$ and $p=5000$ AU spectra. In the $p=10000$ and $p=15000$
AU spectra the observed intensity of the satellite lines is somewhat
stronger than predicted, but its signal-to-noise ratio is poor and
therefore we consider this discrepancy less significant.  Lines are broad
(note that the satellite lines on each side of the main line appear
blended), and the model also reproduces quite well the observed line
widths ($\sim 12$ km s$^{-1}$).  Interestingly, the best fit is obtained
with values of the abundance for the cold and hot parts of the envelope
that coincide respectively with the typical values inferred (usually
assuming a constant abundance) from the observations of low mass (cold)  
and high mass (hot) star-forming cores (see $\S$~\ref{am1}), giving
additional support to the goodness of the fit.

We also show in Figure~\ref{fig:gridvariable} two additional cases,
corresponding to a lower value of the abundance in the hot region of the
envelope, $X_{\rm max}=3\times10^{-7}$ (left column), and to a higher
value of the abundance in the cold region, $X_{\rm min}=2\times10^{-7}$
(right column), to illustrate the behavior of the resulting spectra when
varying these parameters. As Figure \ref{fig:gridvariable} illustrates,
the satellite lines are mainly sensitive to variations in $X_{\rm max}$
(left and middle columns), whereas the main line is sensitive to
variations in $X_{\rm min}$ (middle and right columns).  At large values
of $p$ ($p$ = 10000 and 15000 AU), the spectra do not change by varying
$X_{\rm max}$ because the temperatures along these lines of sight are
lower than 100 K and, therefore, sublimation will not occur.  
Figure~\ref{fig:gridvariable} (left column) also shows that an abundance
of $X_{\rm max}=3 \times 10^{-7}$ (ten times lower than in the best fit)  
in the inner regions of the envelope is high enough to reproduce the
observed brightness temperature of the main line, but it is too low to
reproduce the satellite lines. On the other hand, as the right column
shows, an abundance of $X_{\rm min}=2 \times 10^{-7}$ in the outer parts
of the envelope (ten times higher than in the best fit model) is adequate
to reproduce the brightness temperature of the satellite lines, but not
for the main line, whose peak brightness temperature decreases to a value
below the observed one. This is because for such a high value of $X_{\rm
min}$ the main line becomes very optically thick and the observed emission
originates in the outer (and, therefore, colder) regions of the core.

Since the spacing between values of $M_*$ listed in Table~\ref{tbl5}, that
correspond to those given by Schaller et al. (1992), is too coarse,
especially in the high mass range, we explored additional cases obtaining
the relationship between $M_*$ and $L_*$ by interpolation of the tables of
Schaller et al. In this way, we performed a fine tuning of the mass of the
central star and determined the range of values of the parameters that is
consistent with the observational uncertainties of the data. The results
are summarized in Table~\ref{tbl5b}. In summary, the dust continuum SED
and the spatial variation of the ammonia spectra can be explained by a
dense envelope with a central massive star of mass $M_*$ = 20-25
$M_\odot$, undergoing an intense accretion at a rate $\dot M$ = 2-3
$\times~10^{-3}~M_{\odot}~{\rm yr}^{-1}$, and with a total luminosity of
$L_{\rm tot} \simeq 2 \times 10^5~L_{\odot}$.  The mass of the envelope is
$M_{\rm env}$ = 1400-1800 $M_\odot$ and it has a considerable static
region that surrounds the collapsing region. The value of the magnetic
field is $B$ = 5-6 mG. The outermost region has a velocity dispersion of
$\sigma_{\rm SLS}\simeq 4$ km s$^{-1}$, dominated by turbulent motions.  
The temperature of the envelope ranges from $\sim 40$ K (at $r=30,000$ AU)
to $\sim 1200$ K (at $r$ = 130-160 AU). The gas-phase ammonia abundance
inside the envelope is described by a step-like function with a minimum
value of 1-4 $\times 10^{-8}$ in the outer, colder ($<100$ K) region, and
2-4 $\times 10^{-6}$ in the inner, hotter ($>100$ K) region. This
distribution of the gas-phase ammonia abundance results from the
sublimation of ammonia molecules trapped in water ice grain mantles. The
minimum and maximum values of the ammonia abundance are similar to the
typical values reported in low and high mass protostellar cores,
respectively.


Our results for G31 HMC can be compared with the calculations for this
source made by other authors. Cesaroni et al. (1998) carried out an
analysis of the NH$_3$(4,4) main line assuming local thermodynamic
equilibrium and that the line is optically thick. They claimed that the
observed gradient in the peak brightness temperature is a good estimate of
the radial distribution of the kinetic temperature of the gas inside the
core. Since the observed peak brightness temperature profile of G31 HMC
does not coincide with the expected kinetic temperature profile of a
sphere, these authors concluded that neither a collapsing nor an expanding
spherical envelope can explain the properties of the observed emission.
This conclusion was based on a large velocity gradient (LVG) hypothesis,
that implies that for any line of sight with $p\ne0$ the only contribution
to the peak brightness temperature (at zero velocity relative to the
ambient cloud) arises from the plane perpendicular to the line of sight
that contains the center of the core (since the line-of-sight component of
the infall velocity of this material is strictly zero). However, if lines
are broad (as is the case in G31 HMC), the LVG hypothesis is not valid
since material from outside this plane can also contribute to the emission
at the center of the line.  Furthermore, if the envelope has a static
region (as suggested by our analysis) the peak brightness temperature
would have a significant contribution from the material in this static
region. Therefore, this analysis does not seem adequate to infer the
radial temperature distribution in the envelope and, consequently, the
possibility of a collapsing spherical envelope cannot be discarded.

Cesaroni et al. (1998) favored a geometrically thin disk seen almost face
on (therefore, with a single temperature for each line of sight) to
account for the G31 HMC NH$_3$(4,4) emission. However, in order to reach a
very high optical depth with such a geometrically thin disk a very high
surface density would be required.  This hypothesis was not tested by
Cesaroni et al. (1998), and it is unclear if could correspond to a
realistic scenario. Also, the satellite lines were not considered in the
analysis, and it is also unclear how the observed variation of the main
line to satellites ratio as a function of impact parameter can be
reproduced.

Therefore, we conclude that although we followed a simplified approach (a
spherically symmetric model with a single central source of luminosity),
ignoring deviations from spherical symmetry produced either by rotation
(disks) or due to the intrinsic elongation of the cloud, our model
accounts for some of the basic observed properties of G31 HMC, especially
the highest angular resolution line observations, and constitutes a step
ahead in modeling these structures.

\subsection{Other Ammonia Transitions}
\label{otrastran}

To further test our modeling, we calculated the emerging emission for
ammonia transitions other than the (4,4) using the same SLS model and
ammonia abundances derived by fitting both the SED and the subarcsecond
resolution ammonia (4,4) transition data (Table~\ref{tbl5b}).
Unfortunately, for ammonia transitions other than the (4,4) only
single-dish data (Churchwell et al. 1990, Cesaroni et al.  1992) are
available in the literature. These observations were obtained with the 100
m telescope, with an angular resolution of $40''$. Therefore, a
convolution of the model results with a Gaussian with a FWHM=$40''$ was
performed. Since this angular resolution is insufficient to resolve the
emission of the core, only the spectra towards the $p=0$ position have
been calculated. A value of $V_c=97.4$ km s$^{-1}$ has been adopted for
the LSR velocity of the ambient cloud.

In Figure~\ref{figbaja} we show the model spectra for the (1,1), (2,2),
(4,4), and (5,5) transitions overlaid on the spectra observed by
Churchwell et al.  (1990) and Cesaroni et al.  (1992). The uncertainty in
the absolute calibration of the observed spectra is estimated to be about
30\% (Churchwell et al. 1990, Cesaroni et al. 1992). As can be seen in the
figure, the model spectrum of the (4,4)  line coincides pretty well
(within $\sim$10\%) with the observed one. For the (2,2) and (5,5) lines,
the brightness temperatures of the model spectra are $\sim30$\% lower than
in the observed spectra, which is still within the calibration
uncertainties.  The ratio of intensities of the main to satellite lines,
which is independent of the calibration, is $\sim3.5 \pm 0.9$ for the
observed (2,2) spectrum (the uncertainty in the ratio is estimated from
the rms of the spectrum), in agreement with the value of 3.9 obtained for
the model spectrum. The ratio is $\sim3 \pm 0.8$ for the observed (5,5)  
spectrum, which is also in agreement with the value of 3.5 obtained for
the model spectrum. The observed line width (FWHM) of the (2,2) main line
is $\sim6$ km s$^{-1}$, somewhat smaller than the value of $\sim10$ km
s$^{-1}$ obtained for the model spectrum. Finally, the observed line width
of the (5,5) main line is $\sim9$ km s$^{-1}$, which is very similar to
the value of $\sim10$ km s$^{-1}$ predicted by the model. Given the
coincidence of line ratios and line widths, and taking into account the
uncertainty in the calibration, we consider that the model predictions are
roughly consistent with the observations of the (2,2), (4,4), and (5,5)
transitions. However, uncertainties in the calibration are large (30\%)
and a definitive conclusion is difficult to attain with the present data.
For the ammonia (1,1) line, the brightness temperature of the model
spectrum is significantly lower than that of the observed spectrum, and
the shape of the spectrum is also different. We believe that for the
ammonia (1,1) transition, that traces the colder gas, there is likely a
significant contribution of cold molecular gas from outside the core (see
Hatchell et al. 2000), which is not considered in our modeling. Also, the 
assumption $T_{\rm rot}$=$T_k$ used in deriving our ammonia spectra could 
affect these results.

We conclude that our model parameters, obtained by fitting the dust and
high angular resolution NH$_3$(4,4) transition data, can also reproduce,
within the observational uncertainties, the intensities, line profiles,
and main to satellite line ratios of the multitransitional ammonia data.
Our modeling is carried out by fitting essentially four parameters,
namely, the mass accretion rate and the mass of the central star (or,
alternatively, the stellar luminosity) in a physically self-consistent
collapse model of a SLS envelope, and the minimum and maximum gas-phase
ammonia abundances inside this envelope. We conclude therefore that in
massive protostars, where an important temperature gradient is expected to
be present, an analysis in terms of a realistic, self-consistent model
that takes into account the gradients in the physical properties inside
the core is necessary to properly interpret the observational data.

The modeling we carried out for G31 HMC, including the continuum SED as 
well as the high angular resolution properties of the molecular emission, 
is, to our knowledge, one of the more complete ever made for a hot 
molecular core. This source appears to be a good laboratory to test the 
very early stages of massive star formation.

\section{Summary and Conclusions}

Our main results can be summarized as follows:

\begin{enumerate}

\item We have succeeded in developing a self-consistent model of the 
physical structure of a hot core, described as a spherically symmetric SLS 
envelope infalling onto a massive star. The model predicts the continuum 
emission SED as well as the shape and intensity of ammonia lines.

\item This model is able to reproduce the general observed properties of
G31 HMC. In addition to the SED, it can reproduce the strong intensity of
the main and satellite ammonia (4,4) lines, as well as their spatial
variations as observed with subarcsecond angular resolution. The same
model can roughly reproduce, within the observational uncertainties, the
single-dish ammonia (2,2), (4,4), and (5,5) data. The ammonia (1,1) data
likely have a significant contribution of cold gas outside the core that
has not been included in the model.

\item Although the SED fitting alone cannot constrain adequately the
parameters of G31 HMC, the simultaneous fitting of the SED and the ammonia
emission allows us to obtain a preliminary estimate of the physical
parameters of this source. Our best fit model for G31 HMC has a central
protostar of 20-25~$M_{\odot}$, with an age of 3-4 $\times 10^4$ yr, a
high mass accretion rate of 2-3 $\times 10^{-3}~M_{\odot}$ yr$^{-1}$, and
a total luminosity of $\sim 2\times 10^5~L_{\odot}$. The magnetic field
is 5-6~mG. The mass of the envelope is large, 1400-1800~$M_{\odot}$, and
its temperature is high, ranging from $\sim 40$ K at the outer radius of
the envelope (30,0000 AU) to $\sim 1200$ K at the inner radius (130-160
AU).

A steep increase of the gas-phase ammonia abundance, from $\sim
2\times 10^{-8}$ in the outer parts of the envelope to $\sim 3\times
10^{-6}$ in the inner parts, is required to reproduce the observed spatial
variations of the high angular resolution ammonia spectra. This is
naturally explained as a result of the release (at a characteristic
temperature of $\sim100$ K) of ammonia molecules trapped in water ice
grain mantles. Therefore, taking into account variations in the gas-phase
molecular abundances because of sublimation and freezing-out of molecules
onto grain surfaces appears to be necessary to properly reproduce the
highest angular resolution observations, and will probably be a key
ingredient in the modeling of future line observations with ALMA.

\item Although the actual scenario for G31 HMC is certainly complex
(binaries, rotation, outflows, ...), our simple model can provide a rough
description of the main features of the star formation process in this
source, revealing that massive protostars appear to be able to excite high
excitation ammonia transitions, not only at circumstellar scales, but also
in the larger scale ($\sim$ 30,000 AU) infalling envelopes.

However, the SED of G31 HMC is still incomplete, lacking high 
angular resolution data at wavelengths shorter than 880 $\mu$m. Also, 
high angular resolution line data are only available for the ammonia (4,4) 
transition, and the calibration uncertainties in the lower angular 
resolution observations of other transitions are still too large to 
properly constrain the models. Higher quality data as well as an improved 
modeling would be necessary to make firm statements on the physical 
properties of this source.

\end{enumerate}

\acknowledgments

G.A. and M.O. acknowledge support from MEC AYA2005-08523-C03-03 and MICINN
AYA2008-06189-C03-01 grants (co-funded with FEDER funds), and from Junta
de Andaluc\'{\i}a. S.L. and P.D. acknowledge support from DGAPA-UNAM and
CONACyT (Mexico). We thank J.M. Girart, F. Wyrowski, and S. Kurtz for
communicating data in advance of publication. We thank the referee for a
thoughtful review that help us to improve the paper. This research made
use of data products from the Midcourse Space Experiment (MSX).

\appendix
\section{Excitation Temperature of the Ammonia Inversion Transitions
}\label{apenTex}

Since the inversion transitions inside a given rotational level of the
ammonia molecule are more frequent than the transitions to other
rotational levels, we approximate any metastable inversion doublet by a
two-level model and we estimate the excitation temperature, $T_{\rm ex}$,
of its inversion transition by considering a detailed balance 
between excitation and deexcitation, resulting
  \begin{equation}
 T_{\rm ex} = \left\{{{k}\over{h\nu_0}} 
 \ln \left[
 {{\left(1+ A_{ul} n^{-1} \gamma_{ul}^{-1}\right) e^{h\nu_0/k T_k} - 1}
 \over
 {(1/2){A_{ul} c^2} h^{-1} \nu_0^{-3} n^{-1} \gamma_{ul}^{-1}
 I_r e^{h\nu_0/k T_k}+1}} 
 + 1 \right]\right\}^{-1},
 \label{eq:dosni2}
 \end{equation}
 where $\nu_0$ is the rest frequency of the inversion transition, $n$ is
the total number density of gas molecules (mainly H$_2$ molecules, which 
dominate the collisions), $I_{r}$ is the intensity of the local radiation 
field, $T_k$ is the kinetic temperature, $\gamma_{ul}$ is the collisional 
deexcitation rate coefficient, and $A_{ul}$ is the Einstein spontaneous 
emission coefficient from upper to lower level of the inversion doublet.

In Table \ref{tbl:inver} we list the frequencies adopted for the inversion 
transitions of the metastable levels (1,1) to (6,6) of the ammonia 
molecule (from Pickett et al. 1998).

The collisional deexcitation rate coefficient (from Ho 1977) is taken as
 \begin{equation}
 \left({\gamma_{ul}}\over{\rm cm^3~s^{-1}}\right)=2.27\times10^{-11} 
 \left({T_k}\over{\rm K}\right)^{1/2},
  \label{eq:cdrc}
 \end{equation}
 for all the inversion transitions, since Sweitzer (1978) and Danby et al.
(1988) show that it does not change significantly from one transition to
another. 

The Einstein spontaneous emission coefficient from upper to lower level of
the inversion doublet is given by
 \begin{equation}
 A_{ul}={64 \pi^4 \nu_0^{3} \over {3 h c^3}} {K^2 \over {J(J+1)}} \mu_d^2,
 \label{eq:cee}
 \end{equation}
 where $\mu_d$ is the electric dipole moment of the ammonia molecule whose
value is $\mu_d=1.476$ D (Cohen \& Poynter 1974). The values of $A_{ul}$
for metastable states with $J\le 6$ are listed in Table~\ref{tbl:inver}.  

As an example, Figure~\ref{texs} shows the excitation temperature for the
ammonia (4,4) inversion transition as a function of the density for a
range of kinetic temperatures from 10 to 1000 K. As can be seen from the
figure, the transition is well thermalized for the density and temperature
ranges of HMCs. A similar result is obtained for transitions (1,1) to (5,5).

\section{Calculation of the Absorption Coefficient of the NH$_3$
Inversion Transitions}\label{apenkappa}

The absorption coefficient corresponding to a transition between two 
levels is given by (Estalella \& Anglada 1999):
 \begin{equation}
 \kappa(V)={c^3  \over 8 \pi \nu_0^3} A_{ul} n_u
 \left(e^{h \nu_0/{k T_{\rm ex}}}-1\right)
 \phi(V),
 \label{eq:opa0}
 \end{equation}
 where 
$n_u$ is the number density of particles in the upper level,
and $\phi(V)$ is the profile function that contains the dependence on the 
line-of sight velocity. The profile function is actually the line-of-sight 
velocity distribution function, and is normalized so that 
$\int_{-\infty}^{+\infty} \phi(V) d V = 1$.

For the inversion transition of the $(J,K)$ rotational level of the 
ammonia molecule, the absorption coefficient is given by
 \begin{equation}
 \kappa(V)={c^3\over 8 \pi \nu_0^3} A_{ul} n_u
 \left(e^{h \nu_0/{k T_{\rm ex}}}-1\right)
 \sum_{i=1}^5 x_i \phi_{i}(V),
 \label{eq:opa120}
 \end{equation}
 where $\nu_0$ is the rest frequency of the main line ($i=3$) of the
inversion transition (the values adopted are listed in Table 
\ref{tbl:inver} for metastable levels up to $J=6$), $A_{ul}$ is obtained 
from equation~(\ref{eq:cee}) and listed in Table \ref{tbl:inver}, $n_u$ is 
the number density of ammonia molecules in the upper level of the $(J,K)$ 
inversion doublet, and $T_{\rm ex}$ is the excitation temperature that 
describes the relative populations of the inversion doublet and is 
calculated using equation~(\ref{eq:dosni2}). Due to the quadrupole 
hyperfine structure of the ammonia inversion doublet, the profile function 
is given by $\phi(V) = \sum_{i=1}^5 x_i \phi_{i}(V)$, where $x_i$ is the 
LTE relative intensity of the {\it i}th hyperfine component given in Table 
\ref{tab:propnh3}, and $\phi_{i}(V)$ its profile function that is assumed 
to be a Gaussian. Since the profile function of the overall inversion 
transition is normalized, $\int_{-\infty}^{+\infty} \sum_{i=1}^5 x_i 
\phi_{i}(V) d V = 1$, and because $\sum_{i=1}^5 x_i = 1$, the profile 
function of each individual hyperfine component should be also normalized, 
so that, $\int_{-\infty}^{+\infty} \phi_{i}(V) d V = 1$.  Therefore,
 \begin{equation}
 \phi_{i} (V) = {1 \overwithdelims () 2 \pi}^{1/2} {1 \over \sigma_V}
 e^{-(V - V_i - V_{c} - V_s)^2/2 \sigma_V^2},
 \label{eq:perfinalc}
 \end{equation}
 where $V_i$ is the Doppler velocity corresponding to the frequency shift
of each hyperfine component with respect to the main line ($V_3=0$), 
$V_{c}$ is the line-of-sight LSR velocity of the ambient cloud, $V_s$ is 
the line-of-sight component of the systematic velocity field, and 
$\sigma_V$ is the local dispersion of the distribution of line-of-sight 
velocities due to turbulent and thermal motions.

The density of molecules in the upper level of the inversion doublet, that 
is required in equation~(\ref{eq:opa120}), can be written in terms of the 
density of molecules in the $(J,K)$ rotational state, $n_{JK}= n_u + 
n_l$, using the Boltzmann equation and the excitation temperature,
 \begin{equation}
 n_u = {n_{JK} \over 1 + e^{h \nu_0/k T_{\rm ex}}}.
\label{eq:nu}
 \end{equation}

On the other hand, the total density of ammonia molecules is given by the 
sum of the densities of all the rotational states,
 \begin{equation}
 n_{\rm NH_3} = \sum^{\infty}_{J'=0} \sum^{J'}_{K'=0} n_{J'K'}.
 \label{eq:sumatot}
 \end{equation}

Using the Boltzmann equation, the density of ammonia molecules in the 
rotational state $(J,K)$ can be related to the density in any other 
rotational state $(J',K')$ through
  \begin{equation}
 {n_{J'K'} \over n_{JK}} = {g_{J'K'} \over g_{JK}} 
 e^{-(E_{J'K'}-E_{JK})/k T_{JK,J'K'}},
 \label{eq:bolz2}
 \end{equation} 
 where $g_{JK}$ and $g_{J'K'}$ are the statistical weights of the two
rotational states, $E_{\rm JK}$ and $E_{J'K'}$ their rotational energies, 
and $T_{JK,J'K'}$ is the corresponding rotational temperature.
The statistical weights are given by the relations
 \begin{equation}
g_{JK}=\left\{\begin{array}{ll}
4(2J+1), & K\neq \dot 3 \\
8(2J+1), & K=\dot 3,\ K\neq0 \\
4(2J+1), & K=0,
\label{eq:pesos}
 \end{array}\right.
  \end{equation}
and the rotational energies are given by
 \begin{equation}
 E_{JK}=h[B J(J+1) + (C-B)K^2] ~~~(J=0, 1, 2, \dots; K=0, \pm1, \ldots, \pm J),
 \label{eq:energy}
 \end{equation}
 where $B=2.98 \times 10^{11}$ Hz and 
$C=1.89 \times 10^{11}$ Hz are the rotational constants of the ammonia 
molecule (Townes \& Schawlow 1975).

Solving for $n_{J'K'}$ in equation~(\ref{eq:bolz2}) and substituting in
equation~(\ref{eq:sumatot}) one obtains
 \begin{equation}
n_{\rm NH_3}
= {n_{JK} \over g_{JK}} \sum^{\infty}_{J'=0} \sum^{J'}_{K'=0}
g_{J'K'} e^{-(E_{J'K'}-E_{JK})/k T_{JK,J'K'}}.
 \label{eq:sumatot2}
 \end{equation}

Solving the above equation for $n_{JK}$ and substituting the result in
equation~(\ref{eq:nu}), the population of molecules in the upper level of
the inversion doublet $n_u$ can be obtained in terms of $n_{\rm NH_3}$,
 \begin{equation}
n_u = { n_{\rm NH_3} g_{JK} \over (1 + e^{h \nu_0 /k T_{\rm ex}}) 
\sum^{\infty}_{J'=0} \sum^{J'}_{K'=0}
g_{J'K'} 
e^{-(E_{J'K'}-E_{JK})/k T_{JK,J'K'}}
}.
\label{eq:nutot17nov05}
 \end{equation}

This equation can be finally written in terms of the number density of 
H$_2$ molecules, $n_{\rm H_2}$, using the fractional ammonia abundance, 
$X_{\rm NH_3}=n_{\rm NH_3}/n_{\rm H_2}$. Furthermore, if we assume that 
the relative populations of all the rotational states are characterized by 
the same rotational temperature, that we call $T_{\rm rot}$ (this is 
usually the case in high density regions where collisions dominate and all 
temperatures thermalize to $T_k$), the above equation can be written as
 \begin{equation}
n_u = { n_{\rm H_2} X_{\rm NH_3} g_{JK} e^{-E_{JK}/k T_{\rm rot}} \over
(1 + e^{h \nu_0 /k T_{ex}}) Q(T_{\rm rot})
},
\label{eq:nutot3}
 \end{equation}
where $Q(T)$ is the partition function defined as:
 \begin{equation}
Q(T) = \sum^{\infty}_{J'=0} \sum^{J'}_{K'=0} g_{J'K'} 
e^{-E_{J'K'}/k T}.
\label{eq:particion}
 \end{equation}
 In the calculation of $Q(T)$ usually only metastable rotational states
($J=K$) need to be included, since non-metastable states are short lived
and in general are not significantly populated. The sum is extended up to
a value of $J'$ high enough so that the contribution of higher levels is
negligible.

Substituting equations (\ref{eq:nutot3})  and (\ref{eq:perfinalc}) into
(\ref{eq:opa120}), the expression for the absorption coefficient for the
inversion transition of the $(J,K)$ rotational state of the ammonia
molecule becomes
 \begin{eqnarray}
 \kappa(V) & = & {
 \left(1 \over 128\pi^3\right)^{1/2} {c^3 A_{ul} \over \nu_0^3}
 {n_{\rm H_2} X_{\rm NH_3} \over \sigma_V}
 {e^{h \nu_0/k T_{\rm ex}}-1 \over e^{h \nu_0/k T_{\rm ex}}+1}
 } \nonumber \\
 & & \times {g_{JK} e^{-E_{JK}/k T_{\rm rot}} \over Q(T_{\rm rot})} 
 \sum_{i=1}^5 x_i e^{-(V - V_i - V_{c} - V_s)^2/2\sigma_V^2}.  
 \label{eq:opafinal}
\end{eqnarray}

\section{Condensation and Sublimation of Molecular Species}\label{apeneta}

Let's assume a molecular species whose total amount remains constant,
while the solid versus gas phase ratio, $\eta$, changes as a result of
condensation and sublimation processes due to variations in temperature
and density. Let's further assume that there is a small residual fraction
of molecules that remain in the gas phase even at temperatures well below
the characteristic sublimation temperature, as observations suggest that
the abundance of some molecular species does not drop to zero in the very
low temperature regions. Let's call $X_{\rm min}$ the minimum gas-phase
molecular abundance, reached at low temperatures (where $\eta\gg1$), and
$X_{\rm max}$ the maximum gas-phase abundance, reached in the hot regions
($\eta=0$), where all the molecules are in the gas phase. Therefore,
$X_{\rm mol}$, the gas-phase molecular abundance at a given point will
fulfill the following equation:
 \begin{equation}
 X_{\rm mol}+\eta (X_{\rm mol}-X_{\rm min})=X_{\rm max},
 \end{equation}
 where the first term corresponds to the fraction of molecules in the
gas-phase and the second term corresponds to the fraction in the solid
phase, after correcting for the residual fraction of molecules
(corresponding to $X_{\rm min}$) that do not follow the
condensation/sublimation process.

Therefore, the gas-phase molecular abundance can be estimated as
 \begin{equation}
 X_{\rm mol}={{X_{\rm max} - X_{\rm min}}\over{1+\eta}} +  X_{\rm min}.
\label{eq:laX}
 \end{equation}

Following Sandford \& Allamandola (1993), the ratio, $\eta$, of molecules 
in the solid to gas phase, assuming that only thermal processes are 
involved, can be obtained as
 \begin{equation}
\eta = \left({2 \pi n_{g} a_{\rm g}^2 \over \nu_{v}}\right)
\left({k T_k \over m_{\rm mol}}\right)^{1/2}
\exp(E/k T_k)
\label{eq:laeta}
 \end{equation}
 where $n_{g}$ is the number density of dust grains, $a_{g}$ is the grain
radius, $\nu_{v}$ is the lattice vibrational frequency of the molecule, 
$m_{\rm mol}$ is the mass of the molecule, and $E$ is the binding energy 
of the molecule on the ice surface. For the physical conditions of HMCs 
the exponential term, defined by the binding energy $E$, dominates the 
behavior of this equation. Assuming a given dust-to-gas mass ratio, $R$, 
the density of dust grains can be written in terms of the density, $n$, of 
gas molecules as $n_g=3 \mu_m m_{\rm H} R n/(4\pi a_g^3 \rho_g)$, where 
$\mu_m$ is the mean molecular weight, $m_{\rm H}$ is the mass of the H 
atom, and $\rho_g$ is the density of the material constituting the dust 
grains. Assuming an average radius of the grains $a_g\simeq0.1$ $\mu$m 
(see $\S$~\ref{modeldust}), a ratio $R=1/100$, and a density 
$\rho_g\simeq$ 3 g cm$^{-3}$ (Draine \& Lee 1984), we obtain 
$n_g\simeq3\times10^{-12} n$ and equation~(\ref{eq:laeta}) can be written 
in terms of the gas density.

For the case of sublimation of ammonia ices, substitution of the values of 
$\nu_v=3.45\times10^{12}$ Hz and $E/k=3075$ K (Sandford \& Allamandola 
1993) into equation~(\ref{eq:laeta}) gives
 \begin{equation}
\eta_{\rm NH_3}=5 \times 10^{-30}
\left({n_{\rm H_2} \over {\rm cm^{-3}}}\right)
\left({T_k\over {\rm K}}\right)^{1/2}\exp({\rm 3075~K}/T_k).
 \label{eq:etanh3}
 \end{equation}

 In the case that the ammonia molecules are trapped into water ice, we 
substitute the corresponding numerical constants for water, namely 
$\nu_{\rm v}=2\times10^{12}$ Hz and $E/k=5070$ K (Sandford \& Allamandola 
1993)  into equation~(\ref{eq:laeta}), and the equation becomes
 \begin{equation}
\eta_{\rm H_2O}=8 \times 10^{-30}
\left({n_{\rm H_2} \over {\rm cm^{-3}}}\right)
\left({T_k\over {\rm K}}\right)^{1/2}\exp({\rm 5070~K}/T_k).
 \label{eq:etah2o}
 \end{equation}

\clearpage

\begin{deluxetable}{cccccc}
\tabletypesize{\small}
\tablecaption{Compilation of the Observational Dust Data of G31 HMC 
\label{datos}}
\tablewidth{13.5cm}
\tablehead{
\colhead{}
&\colhead{}
&\colhead{Angular}
&\colhead{Flux}
&\colhead{Aperture} 
&\colhead{}\\
\colhead{$\lambda$}
&\colhead{}
&\colhead{Resolution}
&\colhead{Density}
&\colhead{Size}
&\colhead{}\\
\colhead{($\mu$m)}
&\colhead{Instrument }
&\colhead{($''$)}
&\colhead{(Jy)}
&\colhead{($''$)}
&\colhead{Refs.}
}
\startdata
12   & MSX     & 18   &$2.0\pm0.1$\tablenotemark{a}  & $\sim$ 18 &1\\
12   & IRAS    & 30   &$4.1\pm0.3$\tablenotemark{a} & $\sim$ 30 &2\\
21   & MSX     & 18   &$15.0\pm0.9$\tablenotemark{a}  & $\sim$ 18 &1\\
25   & IRAS    & 30   &$52\pm4$\tablenotemark{a}  & $\sim$ 30 &2\\
60   & IRAS    & 60   &$1090\pm170$\tablenotemark{a}  & $\sim$ 60 &2\\
100  & IRAS    & 120  &$2820\pm400$\tablenotemark{a}  & $\sim$ 120 &2\\
450  & SCUBA   & 9    &$227\pm56$\tablenotemark{a}    & $\sim$ 150 &3\\
850  & SCUBA   & 15   &$55\pm3$\tablenotemark{a}  & $\sim$ 150 &3\\
880  & SMA     & 0.8  &$21\pm4$                       & $\sim$ 8 &4\\
1300 & OVRO    & 3.6  &$4.0\pm0.8$                    & $\sim$ 7 &5\\
1300 & 3m IRTF & 90   &$14.2\pm0.5$\tablenotemark{a}  & $\sim$ 90 &6\\
1350 & SCUBA   & 22   &$4.9\pm1.0$\tablenotemark{a}   & $\sim$ 22 &3\\
1400 & PdBI    & 0.7  &$4.4\pm0.9$                    & $\sim$ 10 &7\\
1400 & BIMA    & 0.5  &$3.6\pm0.7$                    & $\sim$ 3  &8\\
2000 & SCUBA   & 34   &$2.9\pm0.6$\tablenotemark{a}   & $\sim$ 34 &3\\
2700 & PdBI    & 2.1  &$0.7\pm0.2$\tablenotemark{b}   & $\sim$ 5 &9\\
3300 & PdBI    & 1.8  &$0.6\pm0.2$\tablenotemark{b}   & $\sim$ 10 &7\\
3400 & OVRO    & 4.7  &$0.28\pm0.18$\tablenotemark{c} & $\sim$ 10 &5\\
7000 & VLA     & 0.05 &$0.0034\pm0.0005$\tablenotemark{d} & \nodata &10\\
13000& VLA     & 2.2  &$0.05\pm0.01$\tablenotemark{e} & $\sim$ 5  &11\\
\enddata
 \tablenotetext{a}{Taken as an upper limit, because of possible
contamination from nearby sources due to the poor angular resolution
($>5''$) of the observation.}
 \tablenotetext{b}{We applied a rough correction to account for the 
expected free-free contribution at millimeter wavelengths from the nearby HII 
region, estimated from the map shown in Fig. 1e of Cesaroni et al.  1994a.}
 \tablenotetext{c}{The reported value was obtained after a detailed
subtraction of the expected free-free contribution of the nearby HII
region (see Maxia et al.  2001).}
\tablenotetext{d}{Lower limit, because a fraction of the flux density is 
likely resolved out (see $\S 3$).}
\tablenotetext{e}{Upper limit, because the emission at this wavelength is 
probably dominated by free-free emission from the nearby HII region.}
\tablerefs{
 (1) Crowther \& Conti 2003;
 (2) IRAS PSC;
 (3) Hatchell et al. 2000;
 (4) J.M. Girart, priv. comm.;
 (5) Maxia et al. 2001; 
 (6) Chini et al. 1986;
 (7) Beltr\'an et al. 2005;
 (8) F. Wyrowski, priv. comm.;
 (9) Cesaroni et al. 1994b;
 (10) S. Kurtz, priv. comm.;
 (11) Cesaroni et al. 1994a.
}
\end{deluxetable}

\clearpage
\thispagestyle{empty}
\begin{deluxetable}
{l@{\extracolsep{-0.2em}}c@{\extracolsep{-0.2em}}crrccc
c@{\extracolsep{-0.4em}}c@{\extracolsep{-0.7em}}
c@{\extracolsep{-0.3em}}ccccc}
\tabletypesize{\scriptsize}
\rotate
\tablecaption{Parameters Derived from the SED Fitting of 
G31 HMC\tablenotemark{a} 
\label{tbl5}}
\tablewidth{0pt}
\tablehead{
&\colhead{$M_*$\tablenotemark{b}}
&\colhead{${\dot M}$\tablenotemark{c}}
&\colhead{$\beta$\tablenotemark{d}}  
&\colhead{$R_{\rm dust}$\tablenotemark{e}}  
&\colhead{$L_*$\tablenotemark{f}} 
&\colhead{$L_{\rm acc}$\tablenotemark{g}}
&\colhead{$T_{\rm 1000}$\tablenotemark{h}} 
&\colhead{$n_{\rm 1000}$\tablenotemark{i}} 
&\colhead{$V_{\rm 1000}$\tablenotemark{j}} 
&\colhead{$\sigma_{\rm 1000}$\tablenotemark{k}} 
&\colhead{$M_{\rm env}$\tablenotemark{l}}
&\colhead{$t$\tablenotemark{m}} 
&\colhead{$r_{\rm ew}$\tablenotemark{n}} 
&\colhead{$P_0$\tablenotemark{o}} 
&\colhead{$B$\tablenotemark{p}} 
\\
\colhead{Model} 
&\colhead{($M_{\odot}$)} 
&\colhead{($M_{\odot}$\,yr$^{-1}$)}
&
&\colhead{(AU)} 
&\colhead{($L_{\odot}$)} 
&\colhead{($L_{\odot}$)}
&\colhead{(K)} 
&\colhead{(cm$^{-3}$)} 
&\colhead{(km~s$^{-1}$)} 
&\colhead{(km~s$^{-1}$)} 
&\colhead{($M_{\odot}$)} 
&\colhead{(yr)}
&\colhead{(AU)}
&\colhead{(dyn\,cm$^{-2}$)}
&\colhead{(mG)}
}
\startdata

I & 25 & $2.6\times10^{-3}$ & 1.0 & 156 &$79000$ &$1.5\times10^5$ 
&342 &$3.1\times10^{7}$ &4.9 &0.78 &
$1.5\times10^3$ & $3.9\times10^4$ & $2.3\times10^4$ & $2.2\times10^{-6}$
& 5.2 \\

II & 15 & $2.3\times10^{-3}$ & 1.0 & 115 &$21000$ &$7.5\times10^4$ 
&296 &$4.2\times10^{7}$ &3.2 &0.91 &
$2.1\times10^3$ & $2.7\times10^4$ & $1.5\times10^4$ & $4.0\times10^{-6}$
& 7.1 \\

III & 12 & $2.0\times10^{-3}$ & 1.0 & 97 &$10300$ &$5.3\times10^4$ 
& 263 &$4.4\times10^{7}$ &2.7 &0.95&
$2.2\times10^3$ & $2.4\times10^4$ & $1.3\times10^4$ & $4.6\times10^{-6}$
& 7.6 \\

IV & 9 & $1.8\times10^{-3}$ & 1.0 & 82 &$4000$ &3$.6\times10^4$ 
&234 &$4.9\times10^{7}$ &2.1 &1.03 &
$2.6\times10^3$ & $2.0\times10^4$ & $1.1\times10^4$ & $6.1\times10^{-6}$
& 8.7 \\

V & 7 & $1.6\times10^{-3}$ & 1.0 & 70 &$1800$ &$2.5\times10^4$ 
&212 &$5.7\times10^{7}$ &1.7 &1.05 &
$2.8\times10^3$ & $1.7\times10^4$ & $8.8\times10^3$ & $7.4\times10^{-6}$
& 9.7 \\

VI & 5 & $1.5\times10^{-3}$ & 1.0 & 62 &550 &$1.7\times10^4$ 
&194 &$7.3\times10^{7}$ &1.1 &1.20 &
$3.6\times10^3$ & $1.3\times10^4$ & $6.6\times10^3$ & $1.22\times10^{-5}$
& 12.4 \\

VII & 4 & $1.4\times10^{-3}$ & 1.0 & 54 &240 &$1.2\times10^4$ 
&180 &$8.3\times10^{7}$ &0.9 &1.22 &
$3.9\times10^3$ & $1.2\times10^4$ & $5.7\times10^3$ & $1.42\times10^{-5}$
& 13.4 \\

VIII & 3 & $1.2\times10^{-3}$ & 1.0 & 46 &80 &$7.9\times10^3$ 
&163 &$9.6\times10^{7}$ &0.7 &1.28 &
$4.4\times10^3$ & $1.0\times10^4$ & $4.6\times10^3$ & $1.81\times10^{-5}$
& 15.2 \\

IX & 2 & $1.2\times10^{-3}$ & 1.1 & 43 &16 &$5.2\times10^3$ 
&160 &$1.5\times10^{8}$ &0.3 &1.51 &
$6.6\times10^3$ & $6.7\times10^3$ & $3.1\times10^3$ & $3.97\times10^{-5}$
& 22.4 \\

X & 1.5 & $1.1\times10^{-3}$ & 1.1 & 37 &5 &$3.5\times10^3$ 
&145 &$1.8\times10^{8}$ &0.2 &1.60 &
$7.7\times10^3$ & $5.6\times10^3$ & $2.5\times10^3$ & $5.40\times10^{-5}$
& 26.1 \\

XI & 1 & $1.0\times10^{-3}$ & 1.2 & 33 &0.7 &$2.2\times10^3$ 
&140 &$2.6\times10^{8}$ &0.1 &1.84 &
$1.1\times10^4$ & $4.0\times10^3$ & $1.7\times10^3$ &
$1.02\times10^{-4}$ & 35.8
\enddata 
 \tablenotetext{a}{Parameters of SLS models that are consistent with the
observed SED of G31 HMC. Models were obtained adopting a given value of
the mass of the central star, $M_*$, and fitting the observed SED taking
the mass accretion rate, $\dot M$, as the only free parameter. An outer
radius of the envelope $R_{\rm ext}=30000$ AU and a radius of the central
star $R_*=1\times10^{12}$ cm have been assumed in all the models (see 
 $\S$ \ref{structure}). 
 For a given value of $M_*$, the observational uncertainty in the data
points results in a formal uncertainty of about 10\% in the values of
$\dot M$ obtained, and in the remaining derived parameters.}
 \tablenotetext{b}{Mass of the central star. Possible values are in the
range $25~M_\odot > M_* > 1~M_\odot$. The values listed correspond to
those given in the evolutionary tracks of Schaller et al. (1992) in this
range.}
 \tablenotetext{c}{Mass accretion rate.}
 \tablenotetext{d}{Index of the power law that describes the dust
absorption coefficient for $\nu \le 1500$ GHz, inferred from the slope of
the observed SED in the millimeter regime.}
 \tablenotetext{e}{Inner radius of the envelope, calculated as the radius
of the dust destruction front, at the dust sublimation temperature of 1200
K (see procedure in Osorio et al. 1999).} \tablenotetext{f}{Luminosity of
the central star obtained from the stellar mass using the evolutionary
tracks of Schaller et al. (1992).}
 \tablenotetext{g}{Accretion luminosity, $L_{\rm acc}=G{\dot M}M_*/R_*$.}
 \tablenotetext{h}{Temperature at radius $r=1000$ AU.}
 \tablenotetext{i}{Number density of gas molecules at radius $r=1000$ AU.}
 \tablenotetext{j}{Infall velocity at radius $r=1000$ AU.}
 \tablenotetext{k}{Dispersion velocity at radius $r=1000$ AU.}
 \tablenotetext{l}{Mass of the envelope, obtained by integration of the 
density distribution.} 
 \tablenotetext{m}{Time elapsed since the onset of collapse.}
 \tablenotetext{n}{Radius of the expansion wave obtained from 
equation~(\ref{eq:rew}) (see $\S$ \ref{structure}).}
 \tablenotetext{o}{Pressure scale of the SLS model, obtained from 
equation~(\ref{eq:p0}) (see $\S$ \ref{structure}).}
 \tablenotetext{p}{Magnetic field, obtained as $B=(4 \pi P_0)^{1/2}$.}
 \end{deluxetable}

\begin{deluxetable}{ll}
\tabletypesize{\small}
\tablecaption{Parameters of the Best Fit Model of G31 HMC\tablenotemark{a} 
\label{tbl5b}}
\tablewidth{8cm}
\tablehead{
\colhead{Parameter}  
&\colhead{Value} 
}
\startdata
$R_{\rm ext}$ & 30000 AU\tablenotemark{b}\\ 
$R_*$ & $1\times10^{12}$ cm\tablenotemark{b}\\ 
$\beta$ & 1.0\tablenotemark{b} \\
$M_*$ & 20-25 $M_{\odot}$ \\
${\dot M}$ & 2-3 $\times10^{-3}$ $M_{\odot}$ yr$^{-1}$\\
$R_{\rm dust}$ & 130-160 AU \\
$L_*$ & 5-8 $\times10^4$ $L_{\odot}$ \\
$L_{\rm acc}$ & 1.0-1.5 $\times10^5$ $L_{\odot}$ \\
$T(R_{\rm ext})$\tablenotemark{c} & 40 K\\
$T(R_{\rm dust})$\tablenotemark{d} & 1200 K\\
$r_{\rm 100\,K}$\tablenotemark{e} & 6000-6500 AU\\
$M_{\rm env}$ & 1400-1800 $M_{\odot}$ \\
$t$ & 3-4 $\times10^4$ yr \\
$r_{\rm ew}$ & 1.9-2.3 $\times10^4$ AU \\
$P_0$ & 2-3 $\times10^{-6}$ dyn cm$^{-2}$ \\
$B$ & 5-6 mG \\ 
$X_{\rm min}$\tablenotemark{f} & 1-4$\times10^{-8}$\\ 
$X_{\rm max}$\tablenotemark{g} & 2-4$\times10^{-6}$\\ 
\enddata 
 \tablenotetext{a}{Summary of physical parameters of the SLS model that
fits the observed SED and the high angular resolution NH$_3$(4,4) data of
Cesaroni et al. (1998).
 }
 \tablenotetext{b}{Adopted value.}
 \tablenotetext{c}{Temperature at the outer radius of the envelope.}
 \tablenotetext{d}{Temperature at the inner radius of the envelope.}
 \tablenotetext{e}{Radius where the temperature reaches a value of 100 
K, the sublimation temperature of H$_2$O ices.}
 \tablenotetext{f}{Ammonia abundance relative to H$_2$ in the outer part 
of the envelope ($r\ga r_{\rm 100~K}$).}
\tablenotetext{g}{Ammonia abundance relative to H$_2$ in the inner part 
of the envelope ($r\la r_{\rm 100~K}$).}
\end{deluxetable}


\clearpage
\begin{deluxetable}{ccc}
\tablewidth{0pt}
\tablecaption{Parameters of Inversion Transitions of the Ammonia 
Molecule \label{tbl:inver}}
\tablehead{
&\colhead{$\nu_0$\tablenotemark{a}} 
&\colhead{$A_{ul}$\tablenotemark{b}} \\
\colhead{($J$,$K$)} 
&\colhead{(GHz)} 
&\colhead{(s$^{-1}$)}
} 
\startdata
(1,1) &23.6944955 &$1.66838\times10^{-7}$ \\
(2,2) &23.7226333 &$2.23246\times10^{-7}$ \\
(3,3) &23.8701292 &$2.55865\times10^{-7}$ \\
(4,4) &24.1394163 &$2.82264\times10^{-7}$ \\
(5,5) &24.5329887 &$3.08642\times10^{-7}$ \\
(6,6) &25.0560250 &$3.38201\times10^{-7}$ 
\enddata
\tablenotetext{a}{From Pickett et al. 1998.}
\tablenotetext{b}{From equation~(\ref{eq:cee})}
\end{deluxetable}

\clearpage
\begin{table}[p]
\scriptsize
\begin{center}
\caption{Electric Quadrupole Hyperfine Structure of the NH$_3$ Molecule
\label{tab:propnh3}}
\begin{tabular}{rlrrrl}
\tableline\tableline
& & $\Delta\nu_i$ &$V_i$ &\\
$i$ &$F_1 \rightarrow F'_1$ & (MHz) &(km s$^{-1}$) &$x_i$ 
& Notes \\
\tableline
\multicolumn{6}{c}{(1,1)}\\
\tableline
1&$0 \rightarrow 1$ &1.531 &$-19.37$ & 0.11111& external satellite\\
2&$2 \rightarrow 1$ &0.613 &$-7.75$ & 0.13889& internal satellite\\
3&1$\rightarrow$1 + 2$\rightarrow$2 &0 &0 &0.08333+0.41667 = 0.50000 &
main line\\
4&$1 \rightarrow 2$ &$-$0.613 &7.75      & 0.13889& internal satellite\\
5&$1 \rightarrow 0$ &$-$1.531 &19.37     & 0.11111& external satellite\\
\tableline
\multicolumn{6}{c}{(2,2)}\\
\tableline
1&$1 \rightarrow 2$ &2.04 &$-$25.78 & 0.05000& external satellite\\
2&$3 \rightarrow 2$ &1.31 &$-$16.55   & 0.05185& internal satellite\\
3&1$\rightarrow$1 + 2$\rightarrow$2 + 3$\rightarrow$3 &0 &0
&0.15000+0.23148+0.41481 = 0.79629 & main line \\
4&$2 \rightarrow 3$ &$-$1.31  &16.55  & 0.05185& internal satellite\\
5&$2 \rightarrow 1$ &$-$2.04  &25.78  & 0.05000& external satellite\\
\tableline
\multicolumn{6}{c}{(3,3)}\\
\tableline
1&$2 \rightarrow 3$ &2.30 &$-$28.88 & 0.02645& external satellite\\
2&$4 \rightarrow 3$ &1.71  &$-$21.47     & 0.02678 & internal satellite\\
3&2$\rightarrow$2 + 3$\rightarrow$3 + 4$\rightarrow$4 &0  &0
&0.21164+0.28009+0.40178 = 0.89352 & main line\\
4&$3 \rightarrow 4$ &$-$1.71  & 21.47      & 0.02678 & internal satellite\\
5&$3 \rightarrow 2$ &$-$2.30  &28.88      & 0.02645& external satellite\\
\tableline
\multicolumn{6}{c}{(4,4)} \\
\tableline
1&$3 \rightarrow 4$ &2.45 &$-$30.43& 0.01620 & external satellite\\
2&$5 \rightarrow 4$ &1.95  &$-$24.21 & 0.01629& internal satellite\\
3&3$\rightarrow$3 + 4$\rightarrow$4 + 5$\rightarrow$5 &0 &0
&0.24305+0.30083+0.39111 = 0.93500 & main line \\
4&$4 \rightarrow 5$ &$-$1.95  &24.21 & 0.01629& internal satellite\\
5&$4 \rightarrow 3$ &$-$2.45  &30.43 & 0.01620& external satellite\\
\tableline
\multicolumn{6}{c}{(5,5)}\\
\tableline
1&$4 \rightarrow 5$ &2.57 &$-$31.40 & 0.01090& external satellite\\
2&$5 \rightarrow 6$ &2.12  &$-$25.91       & 0.01094& internal satellite\\
3&4$\rightarrow$4 + 5$\rightarrow$5 + 6$\rightarrow$6 &0  &0
&0.26181+0.31148+0.38299 = 0.95629 & main line\\
4&$6 \rightarrow 5$ &$-$2.12  &25.91  &0.01094 & internal satellite\\
5&$5 \rightarrow 4$ &$-$2.57  &31.40  &0.01090 & external satellite\\
\tableline
\multicolumn{6}{c}{(6,6)}\\
\tableline
1&$5 \rightarrow 6$ &2.63 &$-$31.46& 0.00783& external satellite\\
2&$6 \rightarrow 7$ &2.25 &$-$26.92 & 0.00785& internal satellite\\
3&5$\rightarrow$5 + 6$\rightarrow$6 + 7$\rightarrow$7 &0  &0
&0.27422+0.31765+0.37677 = 0.96863 & main line\\
4&$7 \rightarrow 6$ &$-$2.25  &26.92 & 0.00785& internal satellite\\
5&$6 \rightarrow 5$ &$-$2.63  &31.46 & 0.00783& external satellite\\
\tableline\\
\end{tabular}
\end{center}
\end{table}

\clearpage

\clearpage

\begin{figure}
\plotone{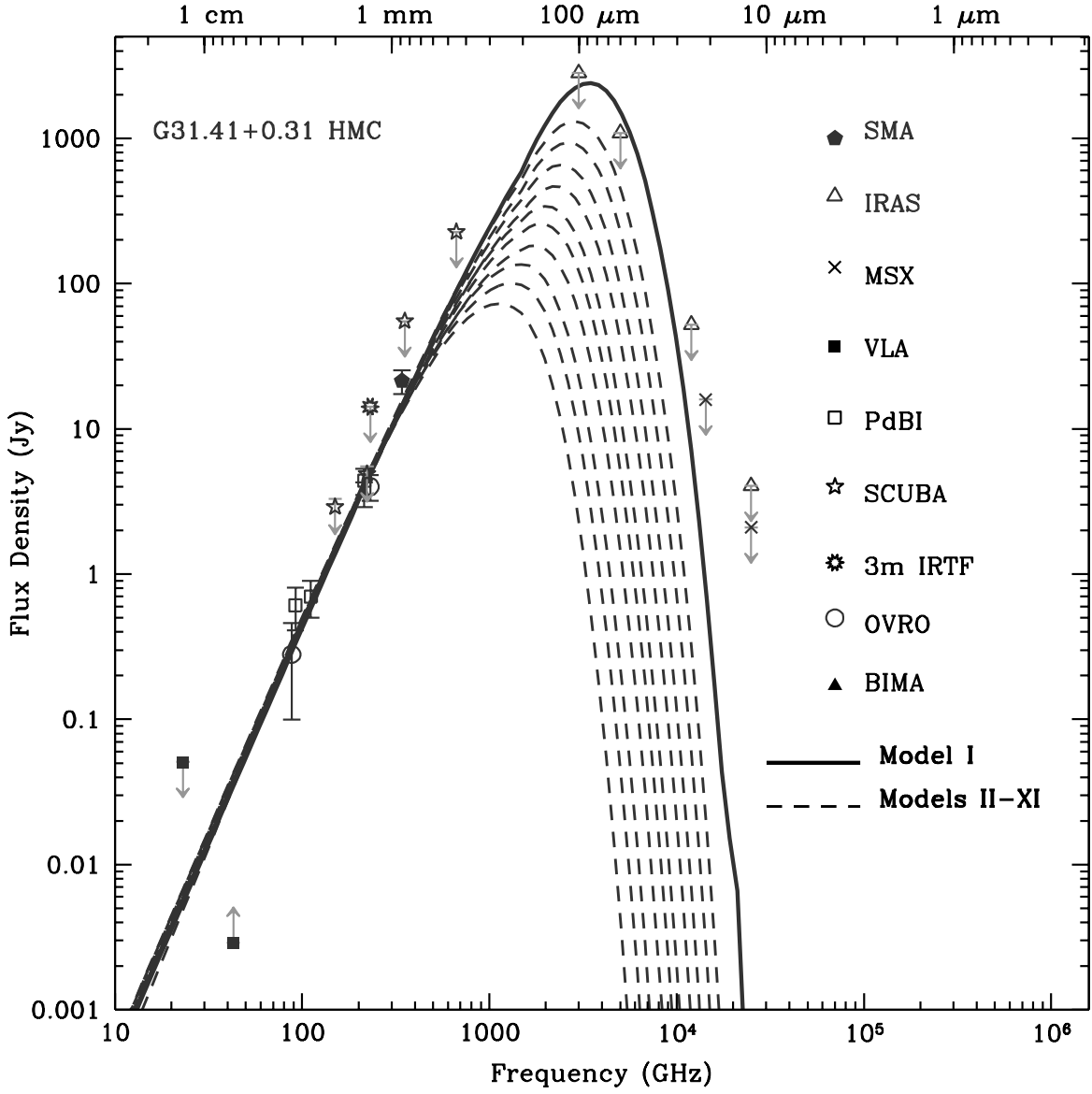}
\caption{Observed flux densities of G31 HMC (see Table \ref{datos}) and 
predicted SED for model I ($M_*=25~M_\odot$; solid line) and models II-XI 
($M_*$ = 15-1~$M_\odot$; dashed lines).}
\label{sed} 
\end{figure}

\clearpage

\begin{figure}
\plotone{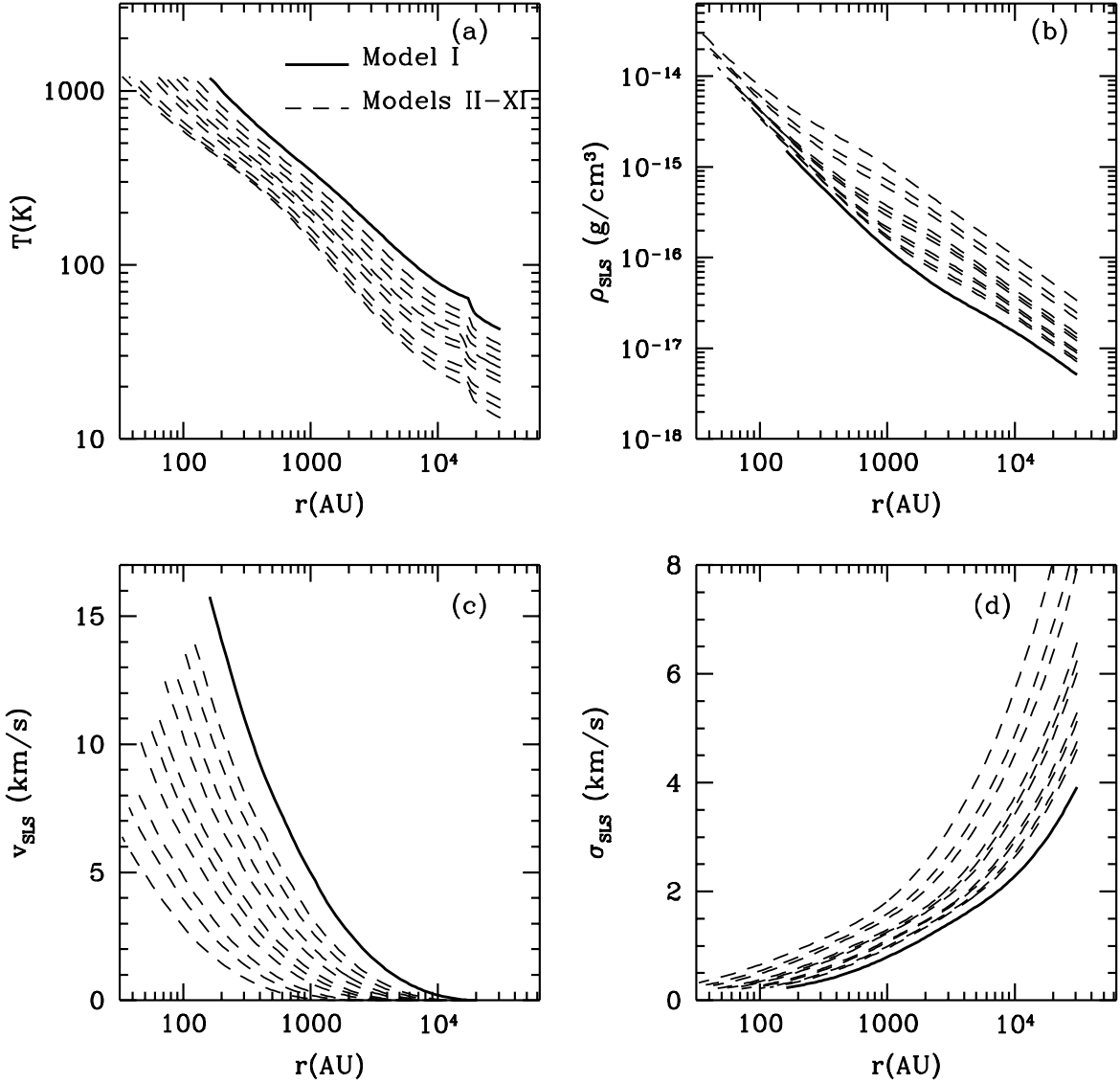}
 \caption{Physical structure of the HMC for model I ($M_*=25~M_\odot$;
solid line) and models II-XI ($M_*$ = 15-1~$M_\odot$; dashed lines).  (a)
Dust temperature as a function of radius; (b) Gas density as a function of
radius; (c) Infall velocity as a function of radius; (d) Turbulent
velocity dispersion as a function of radius.}
 \label{fields}
\end{figure}

\clearpage

\begin{figure}
\vspace{9.0cm}
\includegraphics{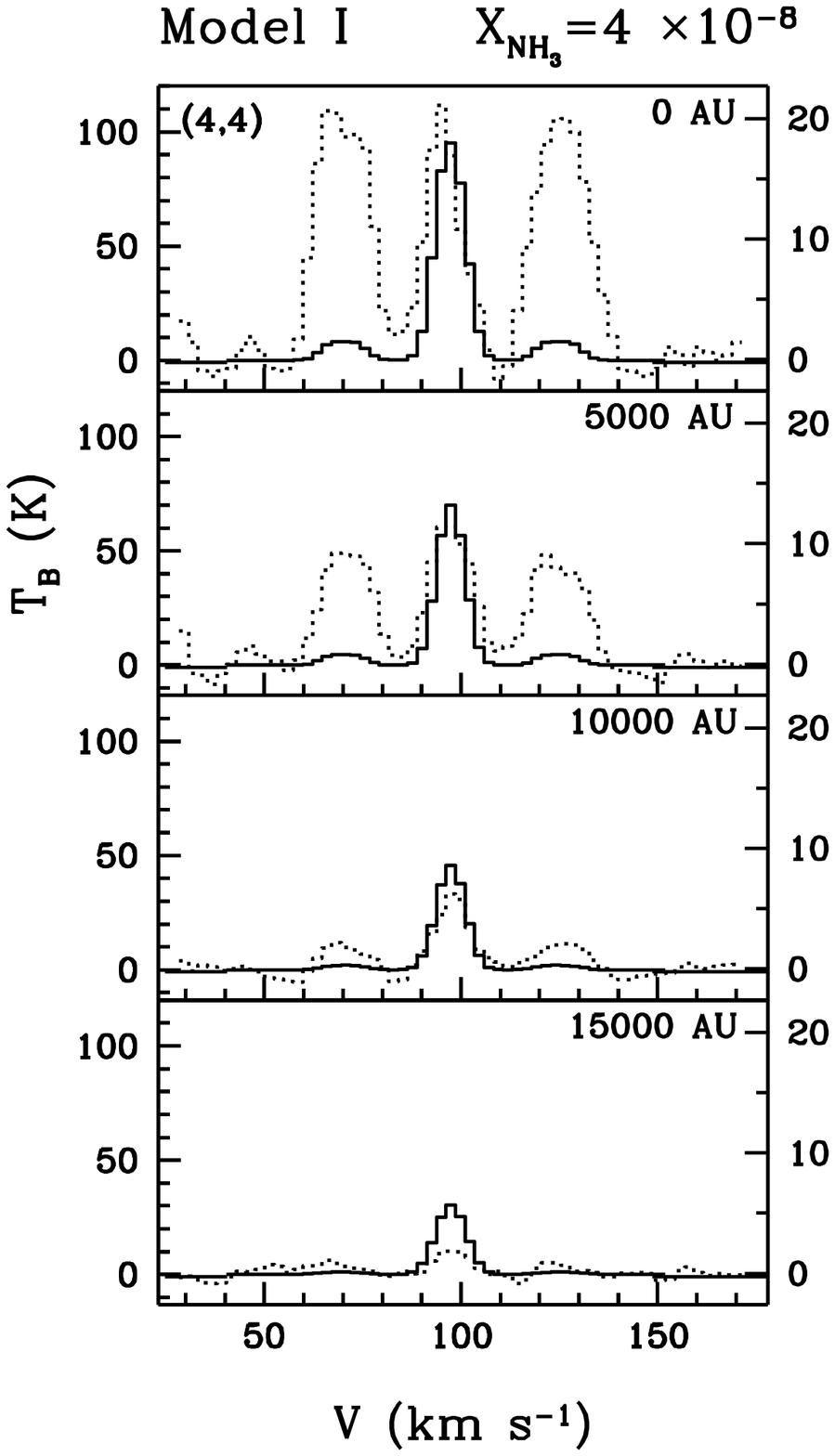}
\includegraphics{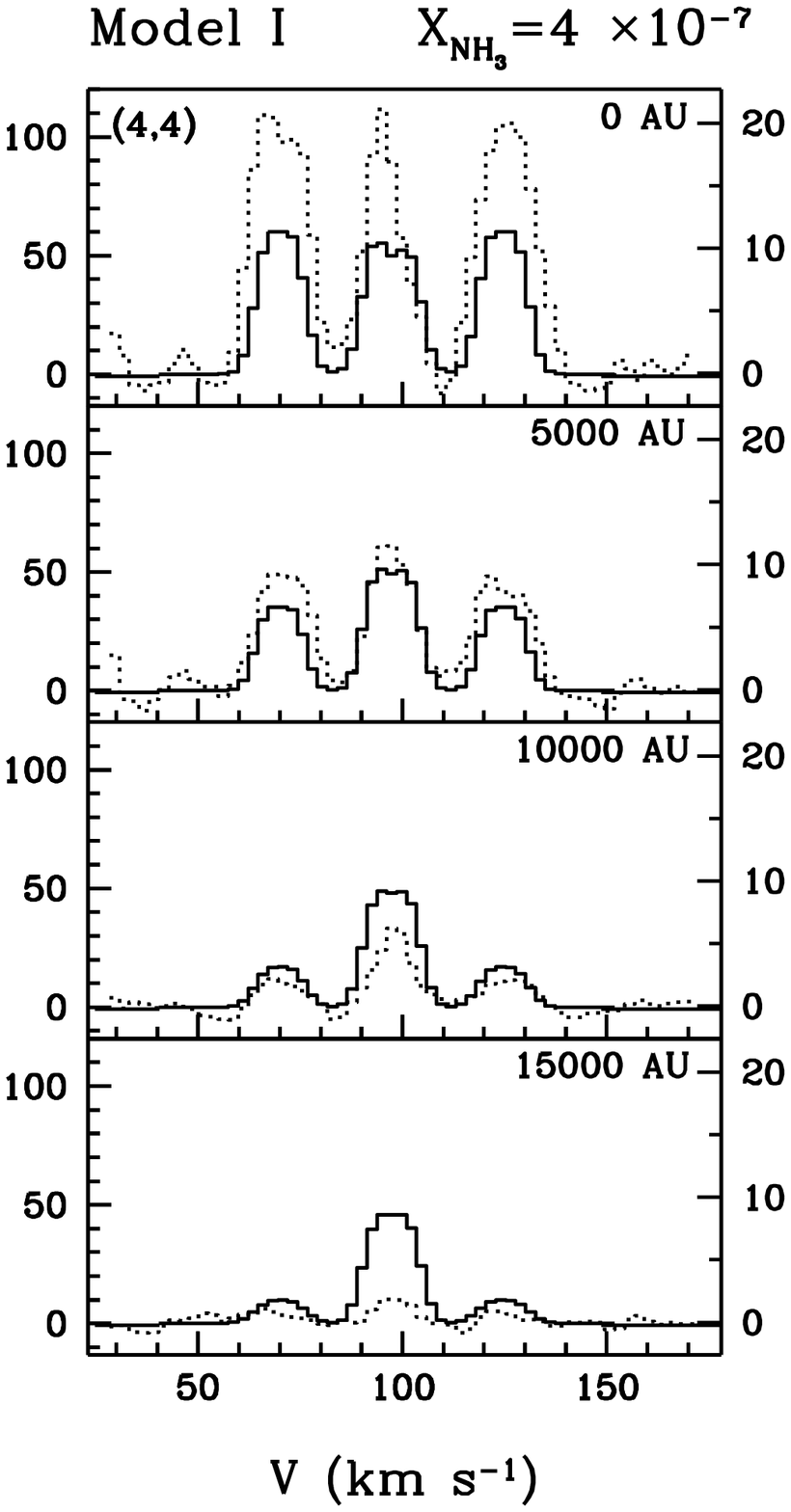}
\includegraphics{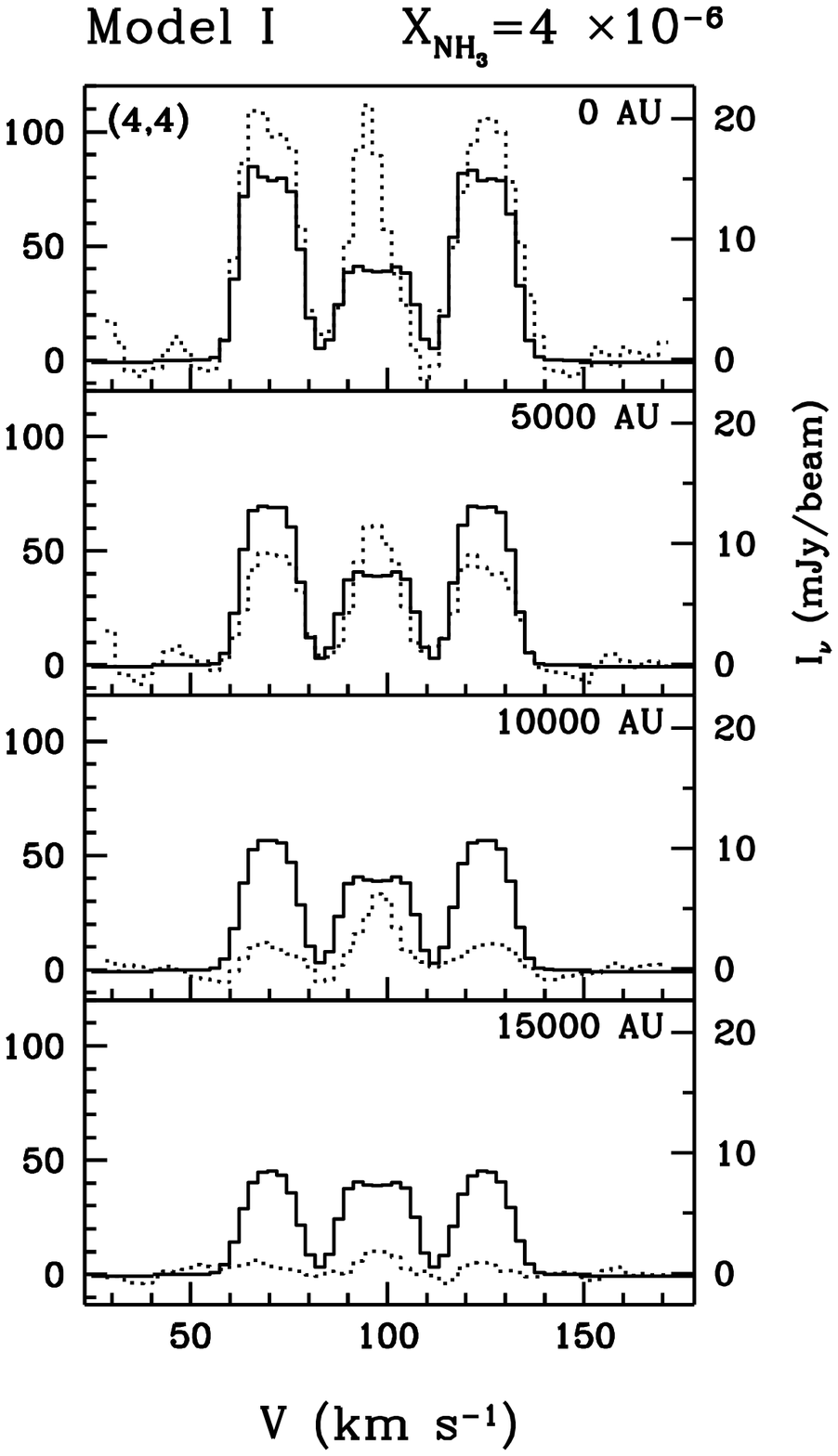}
 \caption{Synthetic spectra of the NH$_3(4,4)$ transition (solid line) for
model I ($M_*=25~M_\odot$) as a function of the impact parameter, for
different values of the gas-phase NH$_3$ abundance, assumed constant along
to the envelope. The values of the abundance are $4\times10^{-8}$ (left
column), $4\times10^{-7}$ (middle column), and $4\times10^{-6}$ (right
column). The values of the impact parameter are indicated in the upper
right corner of each panel.  The observed spectra (adapted from Fig. 9c of
Cesaroni et al.  1998) are plotted in each panel as dotted lines. The
angular resolution is $0\rlap.''63$.}
 \label{fig:grid1} 
\end{figure}

\clearpage

\begin{figure}
\vspace{9.0cm}
\includegraphics{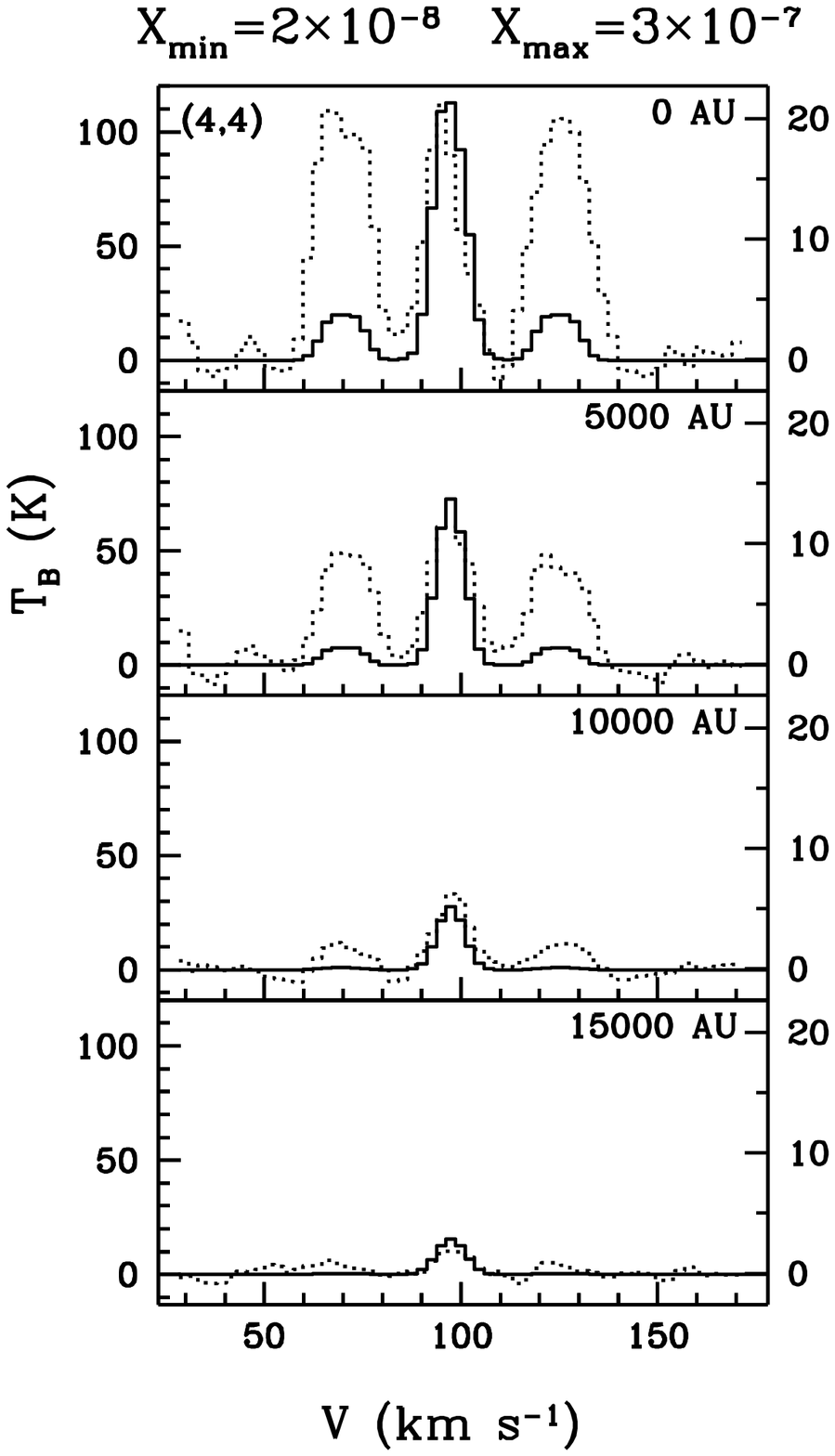}
\includegraphics{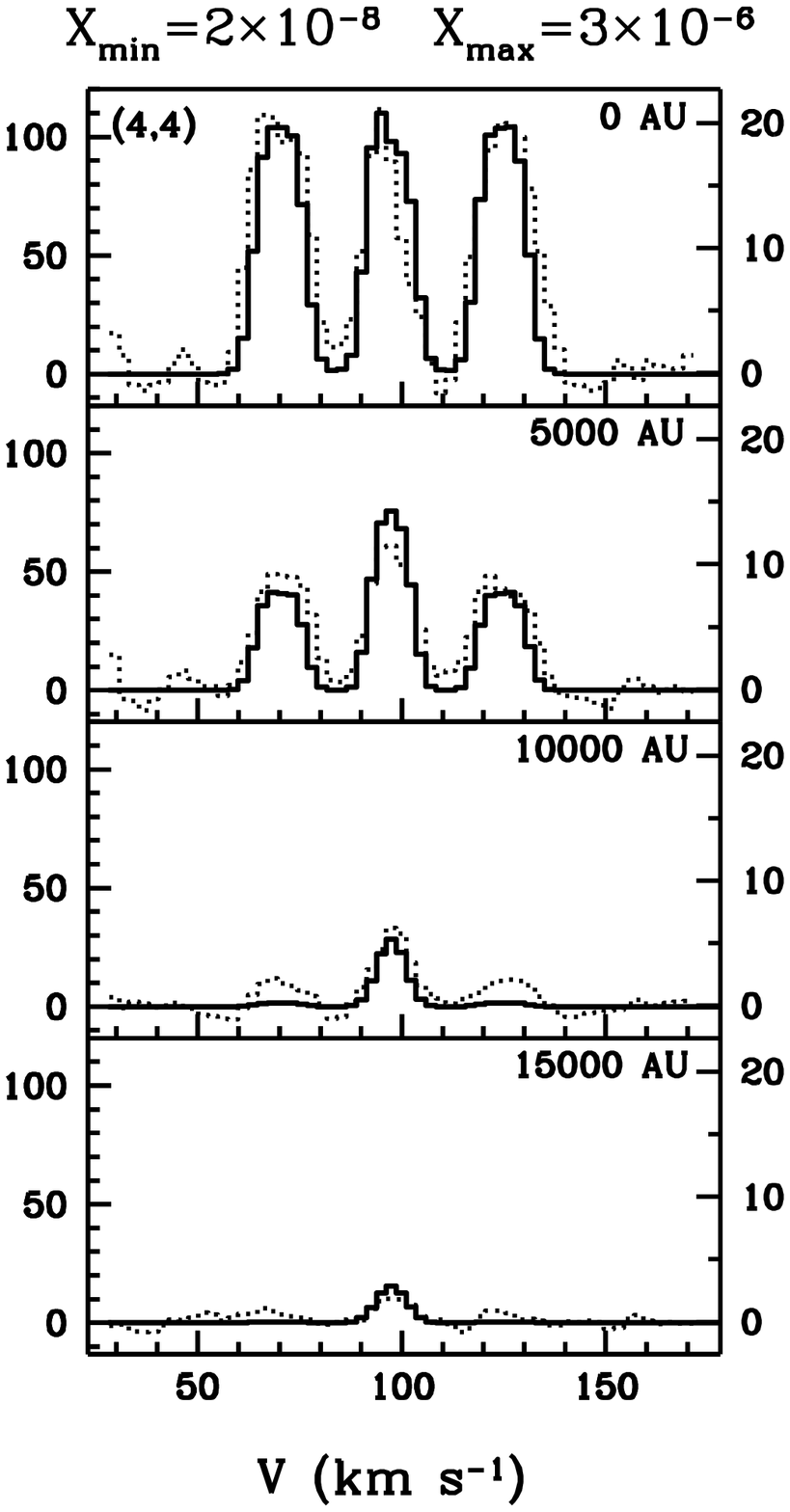}
\includegraphics{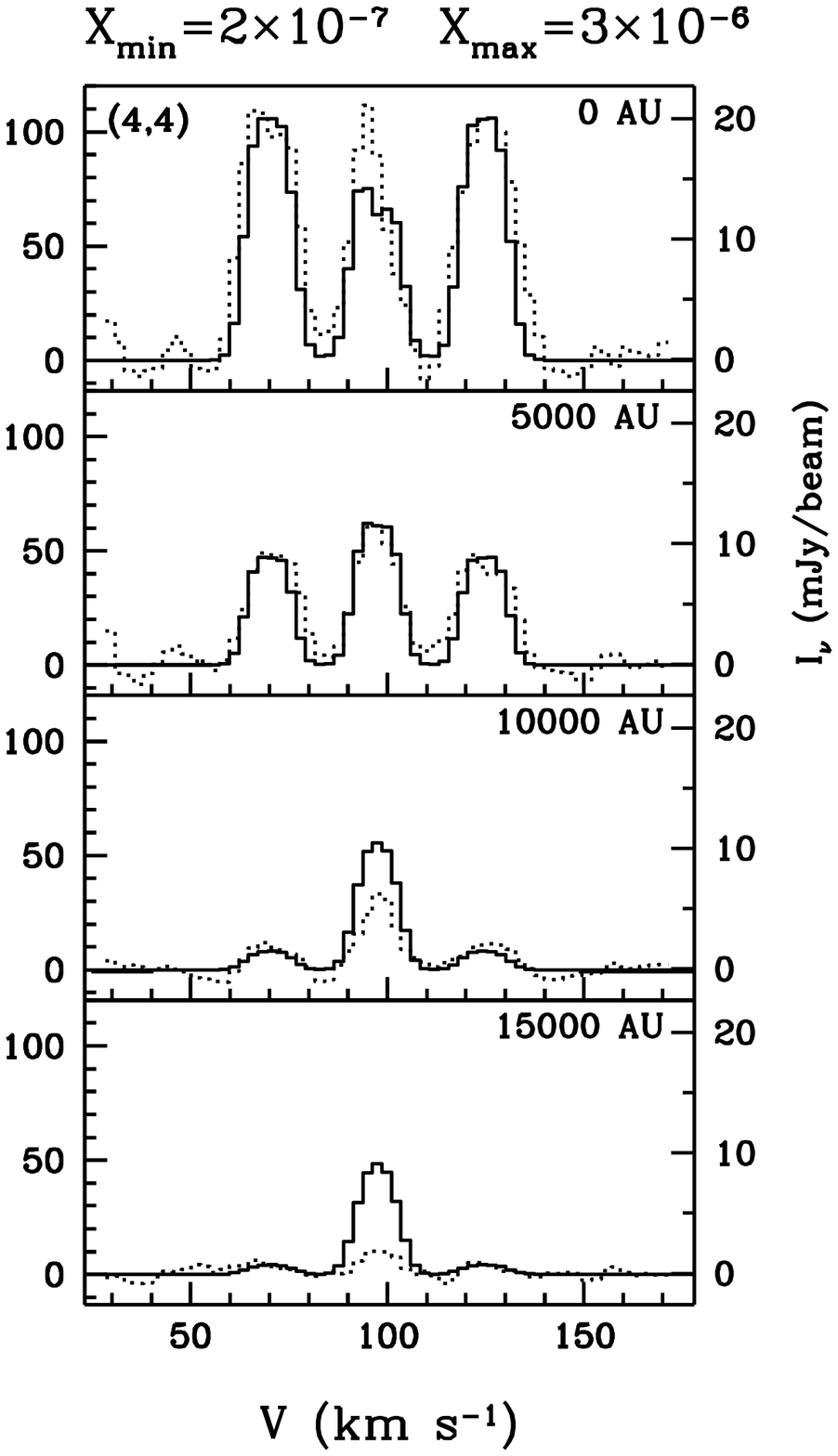}
 \caption{Synthetic spectra of the NH$_3(4,4)$ transition (solid line) for
model I ($M_*=25~M_\odot$) as a function of the impact parameter, assuming
a variable gas-phase NH$_3$ abundance along to the envelope. The values of
the minimum, $X_{\rm min}$, and maximum, $X_{\rm max}$, ammonia abundance
for each case are given on the top of each column. The best fit model,
with $X_{\rm min}=2\times10^{-8}$ and $X_{\rm max}=3\times10^{-6}$, is
shown in the central column. The values of the impact parameter are
indicated in the upper right corner of each panel.  The observed spectra
(adapted from Fig. 9c of Cesaroni et al. 1998) are plotted in each panel
as dotted lines. The angular resolution is $0\rlap.''63$.}
 \label{fig:gridvariable}
\end{figure}

\clearpage

\begin{figure}
\vspace{9.0cm}
\includegraphics{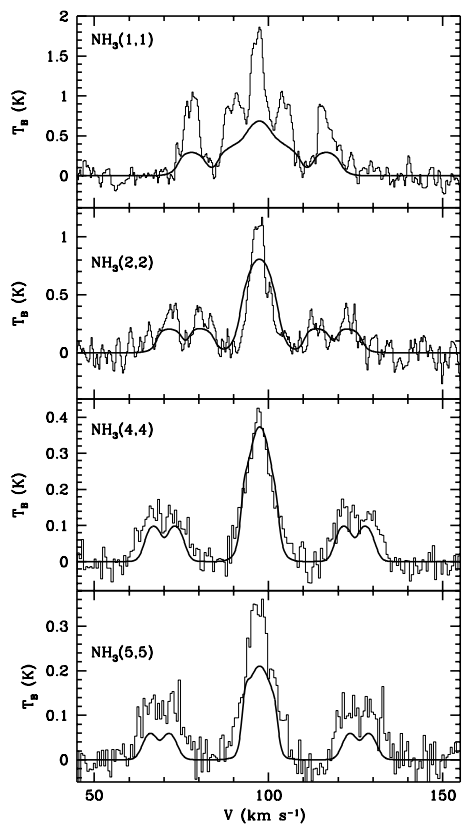}
 \caption{Synthetic spectra (solid line) of the NH$_3(1,1)$, NH$_3(2,2)$,
NH$_3(4,4)$, and NH$_3(5,5)$ transitions towards the center of the HMC,
for an angular resolution of $40''$. The model parameters and the ammonia
abundances are the same as in the best model obtained by fitting the high
angular resolution NH$_3(4,4)$ data (central column in
Fig.~\ref{fig:gridvariable}). The spectra observed with the 100 m
telescope (Churchwell et al. 1990, Cesaroni et al. 1992) are also shown
(dotted lines). The calibration uncertainty in the observed spectra is 
30\%.
}
 \label{figbaja}
\end{figure}

\clearpage

\begin{figure} 
\plotone{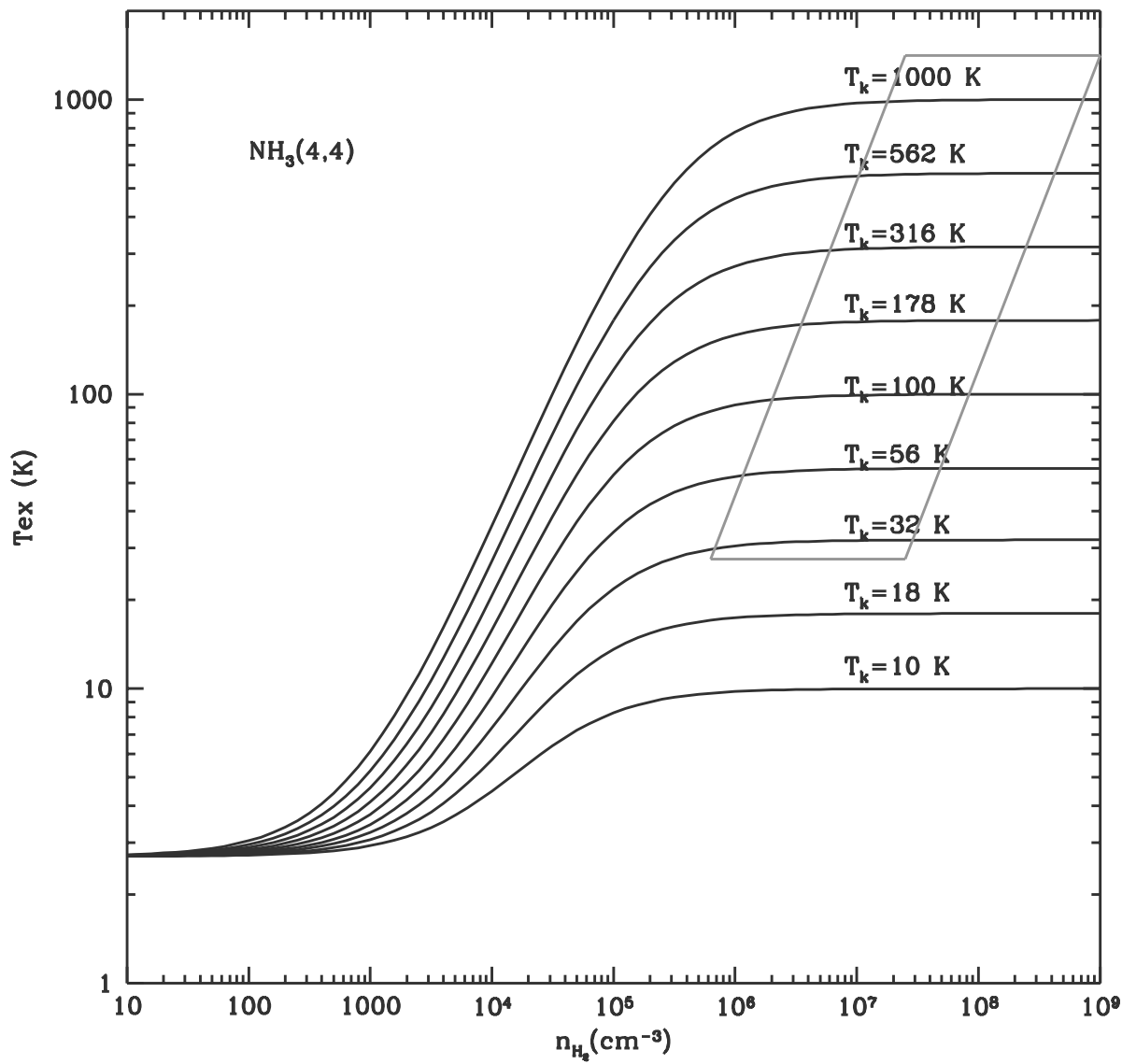}
 \caption{Excitation temperature of the ammonia (4,4) inversion transition
as a function of the molecular gas density, for different values of the
kinetic temperature, obtained from a two-level model. A local radiation
field with the intensity of the cosmological background field, at 2.7 K,
has been assumed. For emission lines, this gives a lower limit to the
excitation temperature. Note that the transition is well thermalized,
$T_{\rm ex}\simeq T_k$, for the range of densities and kinetic 
temperatures of HMCs, which is indicated by a box. }
 \label{texs}
 \end{figure}

\clearpage

\begin{figure}
\plotone{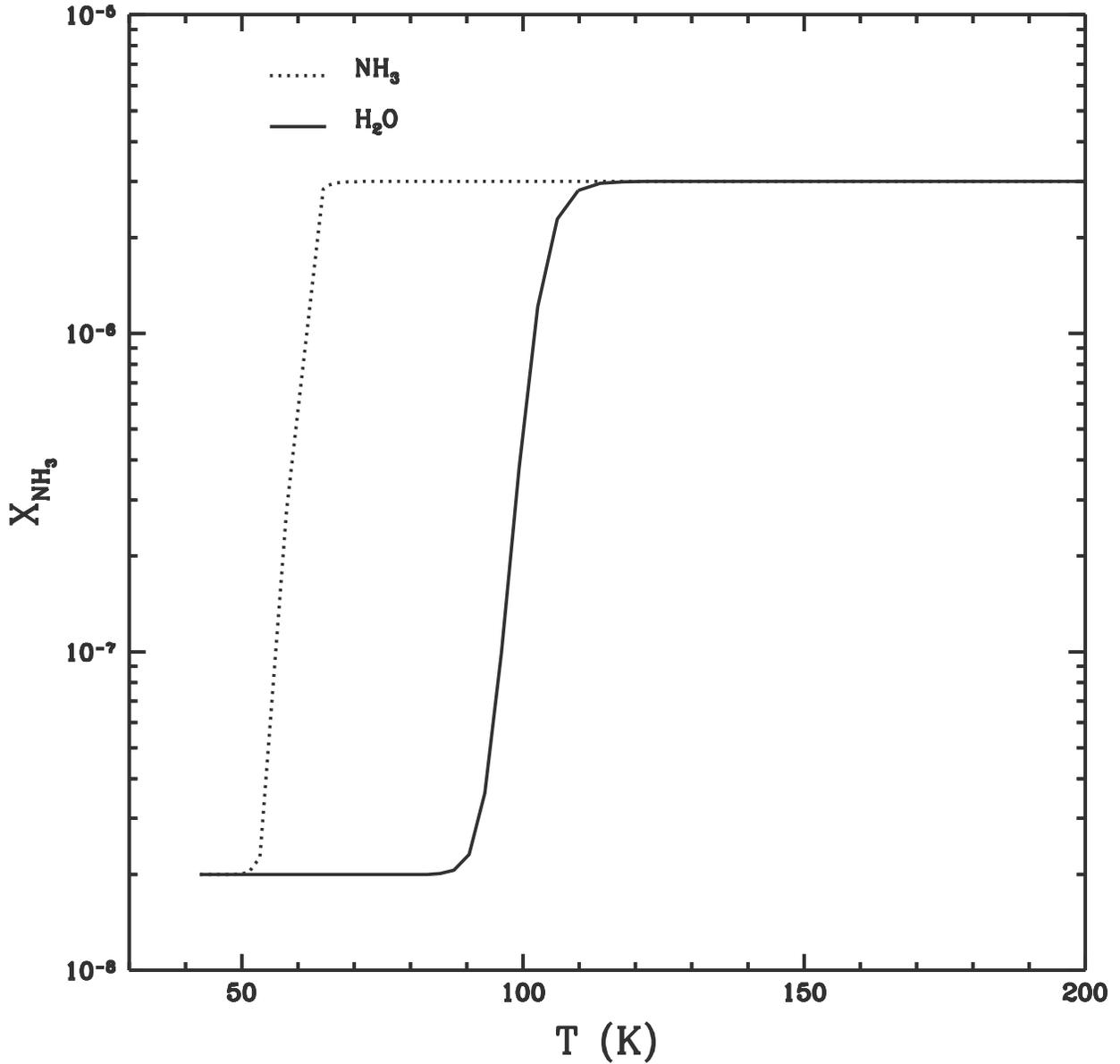}
\caption{Gas-phase ammonia abundance as a function of temperature, 
obtained from the balance between the condensation and sublimation of 
molecules on dust grains following the thermal equilibrium equation of 
Sandford and Allamandola (1993) (see text). The dotted line corresponds 
to the sublimation of pure ammonia molecules, while the solid line 
corresponds to the assumption that ammonia molecules are trapped in water 
ice mantles, being released to the gas phase after sublimation of water 
molecules. For this temperature range, the results depend weakly on the 
density, for which the density distribution field of G31 HMC has been 
assumed. }
 \label{fig:sketch}
 \end{figure}

\end{document}